%% file: main.tex
\documentclass[sigplan,nonacm]{acmart}
\pdfoutput = 1
\usepackage{hyperref}

\usepackage{multirow}
\usepackage{microtype}
\usepackage{graphicx}
\usepackage{amsmath}
\usepackage{amsfonts}
\usepackage{xspace}
\usepackage{footnote}
\usepackage{comment}
\usepackage{subcaption}
\usepackage{mathpartir}
\usepackage{threeparttable}
\usepackage{setspace} 
\usepackage{cleveref}
\usepackage{epsfig}

\usepackage{xcolor}
\usepackage[ruled,algoruled,vlined,linesnumbered]{algorithm2e}
\SetKw{Continue}{continue}
\SetKw{sort}{sort}
\SetKw{descendSort}{descendSort}
\SetKw{key}{key}
\SetKw{Break}{break}
\usepackage[noend]{algpseudocode}
\usepackage{listings}
\usepackage[T1]{fontenc}
\usepackage[scaled=0.78]{DejaVuSansMono}
\usepackage{algorithmicx}
\usepackage{makecell}
\usepackage{colortbl}

\definecolor{mygreen}{rgb}{0,0.6,0}
\definecolor{mygray}{rgb}{0.5,0.5,0.5}
\definecolor{mymauve}{rgb}{0.58,0,0.82}
\long\def\com#1{}

\newcommand{\ie}{{\em i.e.\xspace}}
\newcommand{\eg}{{\em e.g.\xspace}}

\newcommand{\etc}{{\em etc.\xspace}}

\newcommand{\app}{Vela\xspace}

\ifdefined\showRevisions
    \newcommand{\revised}[1]{{\color{blue}#1}}
    \newcommand{\revtag}[1]{\marginpar{\scriptsize\color{blue}\fbox{#1}}}
\else
    \newcommand{\revised}[1]{#1}
    \newcommand{\revtag}[1]{}
\fi

\newcommand{\squishlist}{
   \begin{list}{$\bullet$}
    { \setlength{\itemsep}{0pt}      \setlength{\parsep}{3pt}
      \setlength{\topsep}{3pt}       \setlength{\partopsep}{0pt}
      \setlength{\leftmargin}{3.5mm} \setlength{\labelwidth}{1em}
      \setlength{\labelsep}{0.5em} } }

\newcommand{\squishend}{
    \end{list}  }

\newcommand{\squishnumlist}{
   \begin{list}{$-$}
    { \usecounter{boxlblcounter}
      \setlength{\itemsep}{0pt}      \setlength{\parsep}{3pt}
      \setlength{\topsep}{3pt}       \setlength{\partopsep}{0pt}
      \setlength{\leftmargin}{3.5mm} \setlength{\labelwidth}{1em}
      \setlength{\labelsep}{0.5em} } }

\newcommand{\squishnumend}{
    \end{list}  }

\newcommand{\para}[1]{\smallskip\noindent {\bf #1}}

\setcopyright{none}

\begin{document}
\date{}

\title{Accelerating Stateful Network Applications with Performance Prediction on SoC SmartNICs
}

\author{Shaoke Xi}
\authornote{These authors contributed equally to this work.}
\affiliation{\institution{Alibaba Cloud}\country{China}}

\author{Jiaqi Gao}
\authornotemark[1]
\authornote{Corresponding authors: Jiaqi Gao and Fuliang Li.}
\affiliation{\institution{Alibaba Cloud}\country{China}}

\author{Fuliang Li}
\authornotemark[2]
\affiliation{\institution{Northeastern University}\country{China}}

\author{Minlan Yu}
\affiliation{\institution{Harvard University}\country{USA}}

\author{Ennan Zhai}
\affiliation{\institution{Alibaba Cloud}\country{China}}

\renewcommand{\shortauthors}{Xi et al.}

\makeatletter
\renewcommand{\@mkauthors}{%
  \gdef\@currentauthors{}%
  \gdef\@currentaffiliation{}%
  \global\setbox\mktitle@bx=\vbox{%
    \unvbox\mktitle@bx\par\medskip
    \centering
    {\large Shaoke Xi\textsuperscript{1*}\quad
      Jiaqi Gao\textsuperscript{1*\textdagger}\quad
      Fuliang Li\textsuperscript{2\textdagger}\quad
      Minlan Yu\textsuperscript{3}\quad
      Ennan Zhai\textsuperscript{1}\par}
    {\normalsize\textsuperscript{1}Alibaba Cloud\quad
      \textsuperscript{2}Northeastern University, China\quad
      \textsuperscript{3}Harvard University\par}
    \bigskip}}
\let\vela@authornotes\@authornotes
\renewcommand{\@authornotes}{\setcounter{footnote}{0}\vela@authornotes}
\makeatother

\input{0-abstract}

\ccsdesc[500]{Networks~Programmable networks}
\keywords{SoC SmartNICs, stateful network functions, performance modeling, predictive compilation}

\maketitle
\hypersetup{pdfauthor={Shaoke Xi, Jiaqi Gao, Fuliang Li, Minlan Yu, Ennan Zhai}}

\input{1-intro}
\input{2-motivation}
\input{3-overview}
\input{6_2-Compiling_algorithm}
\input{6_1-Performance_Model}
\input{6-Runtime}
\input{7-Eval.tex}
\input{8-discussion}
\input{10-conclusion}

\newpage

\begin{small}
\bibliographystyle{ACM-Reference-Format}
\bibliography{ref}

\clearpage
\input{appendix}
\end{small}

\end{document}

%% file: 0-abstract.tex
\begin{abstract}
Offloading stateful network functions to multi-threaded SoC SmartNICs promises significant performance and cost benefits.
However, realizing this potential is hindered by two fundamental challenges. 
First, without performance guidance, developers are forced into a slow, manual trial-and-error cycle of deploying and testing to find a feasible resource allocation. 
\revtag{R16}\revised{Second, sustaining performance under changing traffic requires adapting state residency and handling overload within memory layouts fixed at compile time.}
This paper introduces \app, a framework that addresses both challenges through a model-driven, compile-time/runtime co-design. 
Its core is a predictive compiler that replaces the manual tuning loop with fast, automated analysis, 
using a novel, state-centric analytical model to estimate the throughput ceiling of any given resource allocation plan.
This is complemented by a lightweight runtime that \revised{dynamically manages cache contents within the compiled memory layout and mitigates overload.}
We implement and evaluate \app on Netronome Agilio and NVIDIA BlueField 3 SmartNICs \revised{across four NFs}.
\revised{\app reduces host CPU or on-board Arm core usage by 62.9\%--91.9\% relative to the best baseline achieving the same throughput.
For the NATLB workload, \app also improves throughput by 15.2\%--120.8\% over the best baseline at the same host/Arm core count.}
\end{abstract}

%% file: 1-intro.tex
\vspace{-0.5\baselineskip}
\section{Introduction}
\label{sec-intro}

High data center network demands strain traditional CPUs, driving the adoption of SoC SmartNICs 
(e.g., NVIDIA BlueField~\cite{bluefield}, Intel IPUs~\cite{ipu}, AMD Pensando~\cite{pensando}, Netronome~\cite{netronome}) for acceleration. 
Offloading network functions (NFs) to SmartNICs is proven effective~\cite{accelTCP, DBLP:conf/nsdi/KatsikasBKSM18, deepmatch, kvdirect} 
and widely deployed by cloud providers~\cite{aws_nitro, ipu, pensando} to boost performance and reduce total cost of ownership.

Many SoC SmartNICs feature a multi-threaded architecture to balance performance and cost~\cite{deepmatch,accelTCP,shashidhara2022flextoe,siracusano2022re}.
This design matches for the fast path of a NF, which is often already designed for parallel, run-to-completion processing on the host. 
In contrast, complex slow path logic, such as first-packet handling, is ill-suited for the restrictive programming model of these simple parallel threads~\cite{pfaff2015design,tu2021revisiting,pismenny2022nicmem}. 
While this fast-path alignment is a good high-level fit, implementing the offload is still challenging. 
Developers must learn new programming models, hardware interfaces, and performance tuning rules for each specific NIC~\cite{kfoury2024comprehensive,guo2023lognic}.

Significant efforts have been dedicated to simplifying the offloading of fast-path NF logic through compiler design~\cite{floem,gallium,xing2023pipeleon,lyra,flightplan,hao2025p4rtc,lin2025alkali,DBLP:conf/sosp/QiuXHKLNC21}.
By providing high-level programming abstractions, compilers handle the heavy lifting of generating and optimizing NIC code,  
freeing developers from low-level programming and repetitive experience-based performance debugging.
However, current compilers lack a critical capability: they cannot predict the performance of an offload plan before it is deployed.
In real-world scenarios, an operator not only needs to offload a specific set of NFs,  
but also needs to understand the performance consequences of this decision: 
\textit{given a set of NFs and an expected traffic workload, what is the predicted throughput ceiling on the target SmartNIC?}
Without this prediction, operators risk deploying a configuration that overloads the NIC, leading to packet drops. 
Current compilers will dutifully generate code for all given NFs, but cannot warn the operator that the plan might fail under a specific workload. 

A predictive compiler framework changes this workflow. 
It can propose an optimized allocation plan for the expected traffic workload with an associated predicted throughput ceiling.
This enables powerful "what-if" capacity planning: operators can explore trade-offs, determine if a plan is sufficient for peak traffic, 
or use the model's feedback to iteratively refine the set of offloaded NFs. 
Crucially, the prediction also acts as a diagnostic, identifying plans that have hit a fundamental hardware limit where provisioning more resources is the only viable path—all without costly on-device testing. 
\revised{However, deviations from the expected traffic profile require runtime support for cache management and overload mitigation.}

Calculating this predicted throughput ceiling requires modeling the rate at which a workload's demand is met by the SmartNIC's resources.
\revised{NF state is information retained across packets, such as connection mappings and load counters.
We use individual \textbf{state objects}}~\cite{woo2018elastic,pensando_nsdi} as the fundamental analytical unit, which connects software to hardware:
NF logic defines the composition of each state, workload patterns dictate the computational demand on each state, and the hardware architecture determines the resources available to service that demand.
This state-centric view enables a \textbf{compositional} model, allowing us to predict the throughput of any NF combination under any workload by analyzing its constituent states. 
The final throughput is thus dictated by the primary bottleneck that emerges from a given mapping of states to hardware resources.

Our work seeks to unlock the full performance potential of multi-threaded SoC SmartNICs through principled, model-driven resource allocation.
Specifically, there are two paramount dimensions for the state-mapping problem: 
memory placement across the memory hierarchy, and \revtag{R16}\revised{allocation of the SmartNIC's hardware threads to state processing}.

For memory placement, the core challenge is to optimally utilize the scarce, high-performance on-chip SRAM. 
While placing frequently accessed states in SRAM is ideal, SmartNICs' memory layouts are fixed at compile time and cannot be reallocated at runtime~\cite{micro_c}.
This forces a compiler to create a static memory partition optimized for a representative traffic profile, which presents three key problems. 
First, the compiler needs to resolve the NF's logic into its fundamental states and dependencies through deep analysis, and then correlate them with the traffic profile to quantify the resource demand of each state. 
Second, it needs to solve the static allocation problem: 
performing a fine-grained assignment of states to the SRAM to maximize the throughput ceiling for the given profile. 
\revised{Finally, the runtime must manage cache contents within the fixed layout and mitigate overload when live traffic deviates from the original profile.}

For computation allocation, performance does not scale linearly with the number of threads. 
The primary bottleneck arises from atomic state blocks whose multi-instruction operations require software locks to ensure consistency.
On multi-threaded architectures, even small critical sections can cause widespread thread stalls when many threads compete for the same lock,
especially on hardware that lacks strong flow-to-thread affinity. 
Crucially, this lock contention introduces a hard throughput ceiling for the protected states. 
A compiler must therefore be able to predict this ceiling. 
If a state's expected access rate exceeds this limit, the NF is not offloadable; otherwise, the compiler should minimize wasted cycles from contention.

In this paper, we propose Vela, an optimization framework that integrates model-based performance prediction into the compiler workflow for stateful NF offloading.
Vela operates in two phases. 
At compile time (\S\ref{sec:backend}), its Analyzer first derives access patterns from the traffic profile for the identified NF atomic state blocks. 
Then, an iterative, heuristic-driven search begins: 
the Synthesizer proposes resource allocation plans by mapping the finest-grained unit of state (i.e., state element groups) to the NIC's memory hierarchy. 
When a large state object is partitioned across memory tiers, this implicitly creates a software cache in faster memory. 
Each proposed plan is evaluated by our analytical Performance Model (\S\ref{sec-performance}), which predicts its end-to-end throughput. 
The model identifies the primary bottleneck from memory access and, for any states requiring locks, calculates the optimal number of parallel threads and the resulting throughput ceiling.
This loop terminates after exploring the most promising allocation strategies, yielding three key outputs: 
the optimal, hardware-aware allocation plan, its predicted throughput ceiling, and any bottleneck warnings for the operator.
\revised{At runtime (\S\ref{sec:runtime}), \app dynamically manages the compiler-defined software cache and monitors thread idle time to detect overload, redirecting excess traffic to host/Arm cores when needed.}

This paper makes the following contributions:
\squishlist
\item We design and build a model that estimates the throughput ceiling of stateful NFs to quantify the performance impact of both shared memory system bottlenecks and software lock contention among concurrent threads.
\item We design and implement Vela, a system that integrates our model into a practical compiler workflow. 
\app's compile-time synthesizer uses the model's predictions to automatically generate performance-aware resource allocation plans. 
This is complemented by \revised{lightweight runtime caching and overload mitigation.}
\item We demonstrate \app's effectiveness on Netronome Agilio CX and NVIDIA BlueField 3. 
\revtag{R16}\revised{Across four NFs, \app reduces host CPU or on-board Arm core usage by 62.9\%--91.9\% relative to the best baseline achieving the same throughput.
For the NATLB workload, \app also improves throughput by 15.2\%--120.8\% over the best baseline at the same host/Arm core count.}
\app{} is open source at~\cite{vela}.
\squishend

%% file: 2-motivation.tex
\section{Motivation}
\label{sec-motiv}

Offloading network functions to SmartNICs offers significant performance benefits~\cite{kvdirect,deepmatch,fpga_memcached}. 
The primary goal is to \textbf{minimize host CPU usage while meeting throughput demands}. 
However, determining the optimal resource allocation is complex, 
requiring careful consideration of application structure, traffic patterns, and hardware capabilities.

{
\setlength{\abovecaptionskip}{4pt}
\setlength{\belowcaptionskip}{-18pt}
\begin{figure}[t] \centering
\includegraphics[width=\columnwidth]{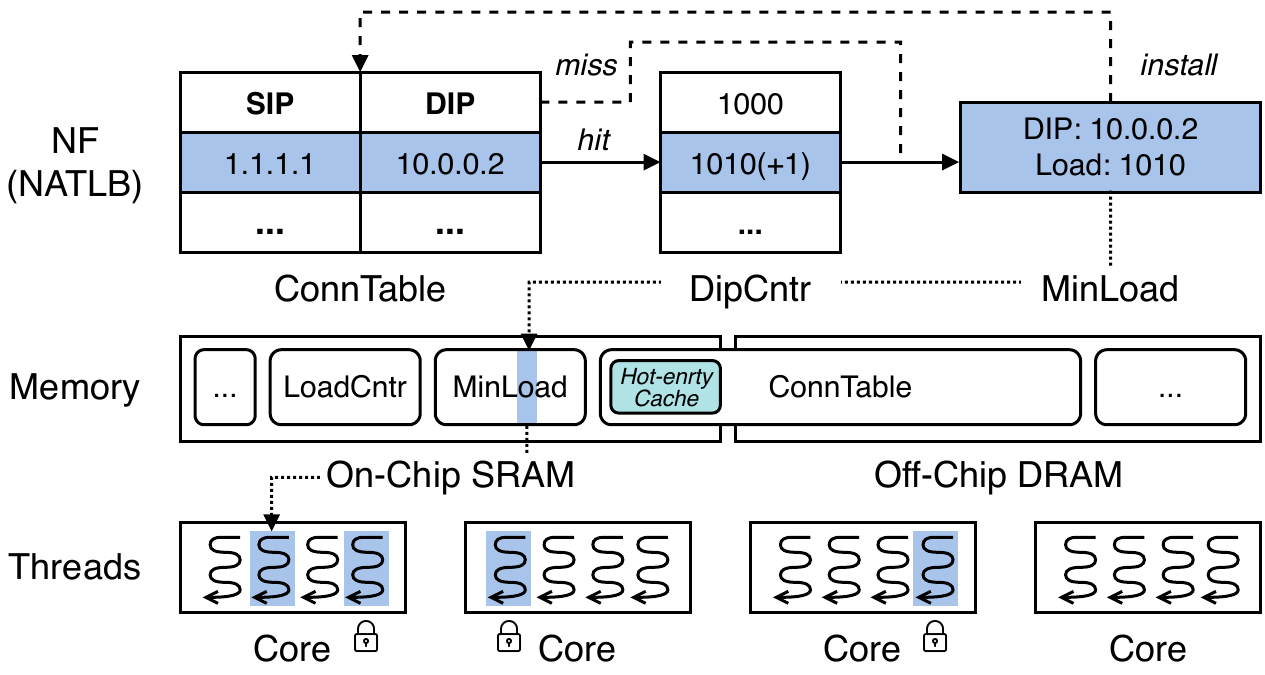}
\caption{State placement across a memory hierarchy and thread contention for shared resources in offloading a NF.}
\label{fig:motiv_example}
\end{figure}
}

\para{Motivating Example.} 
Consider \revised{NATLB, a Layer-4 load balancer that uses network address translation to direct connections to backend servers} (Figure~\ref{fig:motiv_example}).
The NATLB state includes a \textit{ConnTable} (mapping source IPs to chosen backend IPs), 
a \textit{DipCntr} (tracking load per backend), and a \textit{MinLoad} (tracking the least-loaded backend). 
For new connections (a \textit{ConnTable} miss), the control plane queries \textit{MinLoad} and installs a rule in \textit{ConnTable}. 
Then, the first and subsequent packets can hit \textit{ConnTable}, look up the backend DIP, and update \textit{DipCntr} and \textit{MinLoad}, respectively. 
Allocating resources to maximize throughput of executing NATLB on threaded SmartNICs is non-trivial.

\vspace{-0.5\baselineskip}
\subsection{Challenges in Offloading Stateful NFs}
\label{subsec:challenges}

A network application mixes stateless operations (e.g., header checks) and stateful ones. 
Stateless operations are often efficiently handled by NIC hardware if supported; 
our profiling (detailed in Appendix~\ref{sec:appendix:profile}) confirms they rarely form bottlenecks compared to stateful operations. 

The performance of stateful operations varies dramatically. 
\revtag{R7}\revised{
While lookups in hardware-backed match-action \textit{tables} are predictably fast due to specialized acceleration (Appendices~\ref{sec:appendix:profile} \& \ref{appendix:bf2_profiling}),
action-phase accesses to general-purpose, software-defined \textit{state} (e.g., the \textit{DipCntr} array), including software-implemented lookups, map to complex shared memory accesses.}
As shown in Figure~\ref{fig:memArch}, the performance of these accesses hinges on multiple factors: 
the operation type (read vs. write, atomics), the target memory tier (fast on-chip SRAM vs. slower off-chip DRAM), 
and the specific hardware engines involved. 
These factors create a complex, platform-specific performance landscape. 
For example, Agilio employs distinct Atomic vs. Bulk engines for its different memory tiers, 
while BlueField 3 uses a cached memory subsystem that requires a special Memory Aperture path to reach off-chip DRAM. 
The interplay of these variables can lead to 5-10x performance variations for seemingly similar operations (\S\ref{sec-performance}).

{
\begin{figure}[!tp]
\setlength{\abovecaptionskip}{4pt}
\setlength{\belowcaptionskip}{-12pt}
\centering
\includegraphics[width=0.98\columnwidth]{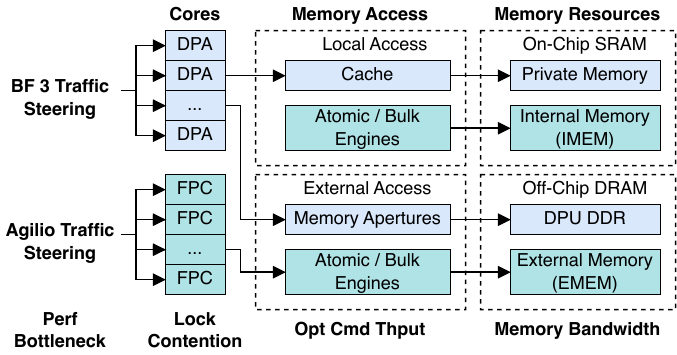}
\caption{Bottlenecks on multi-threaded SoC SmartNICs.}
\label{fig:memArch}
\end{figure}
}

\noindent\textbf{State Placement Optimality.}
This performance complexity creates a difficult optimization problem: 
how to allocate the scarce on-chip SRAM to maximize throughput. 
While the intuitive approach is to place "hot" (frequently-accessed) state in SRAM, 
this simple heuristic is often insufficient.
Different states present complex, competing trade-offs. 
For example, should the compiler dedicate precious SRAM to a small, performance-critical array like \textit{DipCntr} in its entirety, 
or to a software cache holding just the "hot" entries of a much larger \textit{ConnTable}? 
Making the right choice is non-obvious and workload-dependent. 
Answering this requires a model that can quantitatively predict the end-to-end throughput of any given allocation plan.

\noindent\textbf{State Execution Atomicity.}
Synchronization is a core challenge in \revised{multi-threaded} execution.
\revised{
The \textit{MinLoad} update, for example, reads the current load, compares it, and conditionally updates the shared state (code example in Appendix~\ref{appendix:ir}).
On a switching ASIC, such a read-modify-write sequence can execute atomically within a stateful pipeline stage when supported by its hardware primitives~\cite{miao17silkroad}.
On SoC SmartNICs, however, each packet-processing thread executes this sequence as multiple instructions; without a software lock, instructions from different threads can interleave and race on the shared state.}
On parallel hardware, such locks can become severe contention bottlenecks when multiple flows concurrently access the same shared state.

This problem is further shaped by the NIC's hardware dispatching mechanism. 
On BlueField 3, RX steering provides strong flow-to-thread affinity, contention can still arise when distinct flows mapped to the same locked state are steered to different threads. 
By comparison, Agilio's dispatcher does not preserve this affinity: packets are hashed to a cluster of threads, and any idle thread may process them. 
This makes lock-based synchronization essential, even for packets from the same flow.

Lock contention severely impacts run-to-completion performance. 
When a thread is waiting for a lock, it is completely stalled and cannot process any other packets—even those for unrelated flows that do not access the contested state. 
\revised{
Consequently, bursts of accesses to even a single shared state instance can cause widespread thread stalls and limit overall throughput, even when most state accesses are lock-free.}
Therefore, effective offloading requires a model that can quantify the performance impact of lock contention. 
This prediction is essential for determining a locked state's maximum sustainable throughput, 
guiding decisions on whether to offload the state and how to partition threads to mitigate contention bottlenecks.

\vspace{-0.5\baselineskip}
\subsection{Limitation of Existing Frameworks}
\label{subsec:strawman}

{
\setlength{\abovecaptionskip}{4pt}
\setlength{\belowcaptionskip}{-8pt}
\begin{figure*}[t] \centering
\includegraphics[width=\textwidth]{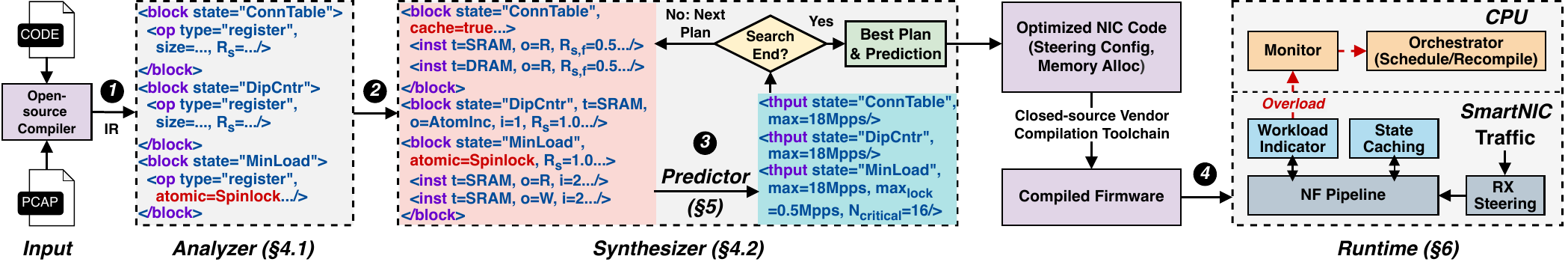}
\caption{\app's architecture and workflow overview.}
\label{fig-arch}
\vspace{-8pt}
\end{figure*}
}

\noindent\textbf{Performance Modeling Approaches.}
While several performance modeling paradigms exist, 
each presents challenges when applied to the specific trade-offs of state placement and resource usage on multi-threaded SmartNICs. 
These approaches can be grouped into three main categories.

\textit{Latency-based Analytical Models}, 
including works like~\cite{DBLP:conf/sosp/QiuXHKLNC21,xing2023pipeleon,lin2025alkali}, 
estimate performance by composing the latencies of primitive hardware operations. 
While computationally efficient for comparing the relative cost of different plans, 
their core assumption—that lower average latency directly yields higher throughput—is misleading on our target hardware. 
On multi-threaded SmartNICs, the latency of individual memory accesses is often hidden by thread parallelism,
breaking the simple correlation between the two metrics.

\textit{Simulation-based models} 
like P4RTC~\cite{hao2025p4rtc} provide high-fidelity throughput estimates but are ill-suited for design space exploration. 
Their primary drawback is a workflow mismatch: they require compiled microcode as input, 
making them prohibitively slow for guiding the iterative, IR-level decisions needed for optimization. 
Furthermore, their reliance on proprietary hardware details makes them viable only with direct vendor support.

\textit{Profiling-based models} 
predict performance from characterization data. 
Black-box approaches like SLOMO~\cite{manousis2020slomo}, for instance, 
can predict contention on shared hardware but cannot map a symptom (e.g., high memory usage) to a specific software cause (e.g., a contended state). 
In contrast, white-box models like LogNIC~\cite{guo2023lognic} use a software execution graph to identify the bottlenecked component.
\app shares this white-box philosophy but adopts a complementary, program-centric perspective. 
Whereas LogNIC's packet-flow model analyzes system-level bottlenecks arising from a program's traversal of hardware components, 
\app's model captures critical intra-component bottlenecks, such as software lock contention, that are dictated by the application's internal state and logic.

\noindent\textbf{Lock Contention Awareness.}
Existing models often overlook software lock contention. 
While capturing hardware contention, they are oblivious to the software-level serialization that can stall numerous threads and trigger performance collapse, leading to unreliable predictions. 
This is crucial, as many stateful NFs, from measurement algorithms~\cite{metwally2005efficient,zhang2021cocosketch} to transport protocols~\cite{li19hpcc,shashidhara2022flextoe,accelTCP}, 
rely on multi-instruction atomic operations that require such locks. 
\app addresses this gap by explicitly modeling lock overhead, 
enabling the compiler to automatically mitigate these bottlenecks (e.g., by partitioning threads) and freeing programmers from complex, error-prone manual redesign.

%% file: 3-overview.tex
\vspace{-1.0\baselineskip}
\section{Overview}
\label{sec-arch}

To address these challenges, we introduce \app, an optimization framework that automates resource allocation for stateful NFs on SoC SmartNICs. 
Integrating into compiler workflows for high-level languages (e.g., P4~\cite{bosshart14p4}, Alkali~\cite{lin2025alkali}), 
\app's compile-time loop and runtime system (Figure~\ref{fig-arch}) analyze NF code against a representative traffic trace to provide predictive performance insights before deployment.

\noindent\textbf{Analysis (\S\ref{subsec:analyser}):}
An open-source compiler frontend parses the NF code into an Intermediate Representation (IR). 
\app's Analyzer then consumes this IR to derive a quantitative model of state access, 
linking the NF's atomic state blocks to the specific request patterns found in the traffic trace.

\noindent\textbf{Synthesis (\S\ref{subsec:synthesizer}) and Prediction (\S\ref{sec-performance}) Loop:}
At the core of compile-time is an iterative loop driven by the Synthesizer. 
It \revtag{R2}\revised{uses a \textit{Global Ordering} heuristic to greedily explore candidate memory placements} for atomic state blocks, including whether to partition a large state object across memory tiers via a software cache.
This plan is then evaluated by the Predictor against the provided traffic profile. 
The Predictor estimates the throughput ceiling by identifying the primary hardware bottleneck that would arise from the specified mix of memory operations.
It also predicts the throughput limit for any locked states and the optimal number of threads to service them. 
This iterative process concludes by selecting the optimal plan for the next stage, 
surfacing its predicted throughput ceiling and any bottleneck warnings (e.g., for states susceptible to lock contention) to the operator as actionable feedback.

\noindent\textbf{Code Generation:}
The open-source compiler backend then uses the chosen plan to generate optimized NIC code, 
embedding the specified memory allocation and thread management constraints, 
which is finally compiled into firmware by the closed-source vendor toolchain.

\noindent\textbf{Runtime Support (\S\ref{sec:runtime}):}
At runtime, \app manages the live application. For tiered state, it activates pre-allocated software caches, dynamically managing hot entries in SRAM.
The deployed NF pipeline is instrumented with a lightweight workload indicator.
When the host-side Monitor detects an overload using this indicator, the Orchestrator first mitigates it as a short-term burst by scheduling excess traffic to other resources (e.g., host). 
For sustained overloads, it triggers a full recompilation with an updated traffic profile to adapt the allocation plan.

%% file: 6_2-Compiling_algorithm.tex
\vspace{-0.5\baselineskip}
\section{\app Compile-Time Framework}
\label{sec:backend}

\app's compile-time framework analyzes a given set of candidate NFs to synthesize an optimized resource allocation plan for a target SmartNIC. 
The goal is to find a plan that maximizes the total computation handled by the NIC. 
\revtag{R1}\revised{We follow the NATLB example from Figure~\ref{fig:motiv_example}: the Analyzer identifies atomic state and its access profile (\S\ref{subsec:analyser}), and the Synthesizer uses these inputs to search memory layouts and generate an executable plan (\S\ref{subsec:synthesizer}).
The example targets Agilio, with IMEM and EMEM as the SRAM and DRAM tiers, respectively.}

\begin{table}[t]
    \centering
    \footnotesize
    \caption{Key notations for the compile-time framework.}
    \vspace{-6pt}
    \label{tab:compile_notations}
    \begin{tabular}{lp{6.5cm}}
    \hline
      \textbf{Symbol} & \textbf{Description} \\ \hline\hline
      $\mathcal{S}$ & Set of state blocks in NFs. \\
      $s \in \mathcal{S}$ & A specific state block. \\
      $\mathcal{T}_s$ & Set of flow group identifiers (indices) for state block $s$. \\
      $f \in \mathcal{T}_s$ & A specific flow group identifier. \\
      $(s, f)$ & A state element group (all state accessed by index $f$ in $s$). \\
      $\mathcal{I}_s$ & Instructions executed when accessing state block $s$. \\
      $R_s$ & Relative execution rate of state block $s$ from traffic trace. \\
      $R_{s,f}$ & Relative access rate of state element group $(s, f)$ from trace. \\
      $\mathcal{A}$ & An allocation plan, mapping each $(s, f)$ to a memory tier. \\
      \hline
    \end{tabular}
    \vspace{-10pt}
\end{table}

\vspace{-0.5\baselineskip}
\subsection{State Analysis with IR Augmentation}
\label{subsec:analyser}

The \textbf{Analyzer} takes NF code and a sample traffic trace as input. 
\revtag{R6}\revised{An open-source compiler frontend, such as p4c or an LLVM-based frontend, lowers the NF code into an intermediate representation (IR) from which standard control-flow graphs (CFG) and data-dependence graphs can be constructed.}
The Analyzer \revised{extracts state access patterns and dependencies from these IR structures~\cite{sivaraman16packet,DBLP:conf/sosp/QiuXHKLNC21,hao2025p4rtc} and} augments the IR with detailed \revised{state-usage metadata} (Table~\ref{tab:compile_notations}), which serves as the formal input to the Synthesizer.

\revised{First, the Analyzer identifies all stateful memory objects and builds a dependency graph among them using reaching definitions analysis on the IR and control dependencies derived from the CFG.}
Objects involved in cyclic read-after-write dependencies are grouped into the same \textbf{atomic state block} ($s \in \mathcal{S}$).
Within a block, the set of state variables accessed together by a single flow group identifier constitutes a state element group.
\revtag{R1}\revised{For NATLB, the Analyzer extracts three blocks: \textit{ConnTable}, \textit{DipCntr}, and \{\textit{MinLoad}, \textit{MinDip}\}.
The last block groups the dependent load and destination-IP updates into one atomic operation.
For a particular server-group index $f=\textit{vip\_id}$, the pair \{\textit{MinLoad}[$f$], \textit{MinDip}[$f$]\} is one state element group $(s,f)$; different server-group indices identify different instances of the same block.
The P4 code and dependency graph are detailed in Appendix~\ref{subsec:frontend:representation}.}
\revised{The Analyzer conservatively groups state objects so that operations requiring joint atomicity remain in the same block.
This preserves state consistency under concurrent execution, although unnecessary grouping may widen critical sections and reduce performance.}
For each block, the Analyzer identifies whether its atomicity can be guaranteed by single hardware atomic instructions or if it requires a software spinlock. 
It also extracts the set of memory-access instructions ($\mathcal{I}_s$) executed when the block is accessed.

Next, it uses the sample traffic trace to derive execution pattern metadata. 
A functional simulation (e.g., using BMv2 for P4) is used to log the code path traversed by each packet, based on which the Analyzer calculates two key metrics:
(1) The relative execution rate for each state block $s$ ($R_s$), which is used by the Predictor to weigh the impact of different instruction mixes when calculating overall throughput.
\revtag{R1}\revised{In the running NATLB example, the access profile visits each block once per packet ($R_s=1$).}
(2) The relative access rate for each state element group ($R_{s,f}$), which reveals "hot" elements for placement optimization.
\revised{In the same example, the hottest quarter of \textit{ConnTable} entries accounts for 80\% of its accesses.}
\revtag{R10}\revised{Collecting these access rates requires the simulator to preserve the NF's control-flow and state-indexing semantics, but not the timing or microarchitectural behavior of the target NIC.
The Predictor separately obtains hardware performance characteristics from primitive performance curves microbenchmarked on the target NIC (\S\ref{subsec:perform:roofline}).}

While a detailed trace allows automatic pattern derivation, \app can also operate with high-level user estimates based on domain knowledge. 
For instance, an operator can specify the relative execution proportions of different state blocks ($R_s$) based on expected traffic splits, 
and use a common statistical model (e.g., Zipf) to synthesize the access distribution ($R_{s,f}$) for state objects that follow known patterns (e.g., key-value stores~\cite{ghigoff2021bmc}).
This enables \app to generate reasonable allocation plans even without available traffic data.

\vspace{-0.5\baselineskip}
\subsection{Synthesis Algorithm}
\label{subsec:synthesizer}

Given a set of candidate NFs, the \textbf{Synthesizer}'s goal is to find an allocation plan $\mathcal{A}$ that maximizes the total offloaded computation. 
We estimate the offloaded computation workload as the product of the plan's predicted throughput $P_{\mathcal{A}}$ and the average computational work per packet.
The average work is \revised{measured by the memory-access instruction count weighted by execution rate}.
This defines our objective function, which we seek to maximize: $P_{\mathcal{A}}\cdot \sum_{s \in \mathcal{S}} (R_s \cdot |\mathcal{I}_s|)$.
\revtag{R2}\revised{Here, $P_{\mathcal{A}}$ denotes the plan's predicted throughput ceiling in the absence of lock contention (\S\ref{subsec:perform:roofline}).
For each candidate layout, the Predictor also derives the thread count and per-instance throughput ceiling for locked state (\S\ref{subsec:perform:lock}).
Thread sizing is thus nested within the memory-layout search.}

\para{Allocation Granularity.} 
The finest unit of allocation is the state element group $(s,f)$.
This entire group is always co-located to avoid complex cross-memory-tier accesses within an atomic operation block.
\revtag{R1}\revised{For example, different NATLB server groups may occupy different tiers, but each \{\textit{MinLoad}[$f$], \textit{MinDip}[$f$]\} pair stays together.}

\para{Hardware Table Acceleration.} 
Before general memory allocation, the Synthesizer first attempts to place explicit match-action table objects onto dedicated hardware accelerators. 
If successful, their performance is governed by a separate, predictable hardware model~\cite{xing2023pipeleon} and they are removed from the main allocation problem.
However, this acceleration may not be feasible for more complex software-defined state, or when the accelerator's limited capacity is exhausted.
For example, a hash table with a custom collision resolution scheme (like cuckoo hashing) requires flexible logic that cannot be mapped to rigid match-action hardware. 
The Synthesizer treats these implicit tables as general-purpose register arrays and manages their placement using its main allocation algorithm.
\revtag{R1}\revised{The running example considers this general-memory path for \textit{ConnTable}, so its placement remains part of the search.}

\para{Software Cache Overhead.}
When partitioning a large state object, a software cache is implicitly created by allocating a region in SRAM to hold "hot" elements that are promoted at runtime from their primary storage locations in DRAM.
The Synthesizer must account for two sources of overhead.
First, it budgets for the SRAM memory cost of the required index remap table.
Second, the Predictor accounts for the lookup overhead,
weighing the cost of an extra SRAM indirection for every access against the significant latency savings of avoiding a full DRAM access for hot elements.
\revtag{R12}\revised{Under uniform or weakly skewed access, the extra lookup may outweigh the savings from SRAM hits.
Small objects that fit entirely in SRAM need no software cache.}
\revised{For larger objects, a} partitioned plan is only considered if both the total SRAM memory consumption (hot entries + remap table) is viable and the predicted performance surpasses that of a DRAM-only allocation.
\revtag{R1}\revised{Our example's \textit{ConnTable} has 131,072 entries of 32 bytes each.
Caching its hottest quarter requires 1\,MiB for the 32,768 cached entries, plus 512\,KiB for a four-byte index per table entry.
This placement serves 80\% of data accesses from SRAM under the sampled profile, with an additional index lookup for every access.}

\begin{algorithm}[tp]
\caption{State Allocation Algorithm.}
\label{alg:allocation}
\begin{footnotesize}
\KwIn{Set of all candidate NFs, $\mathcal{N}_{all}$, and their associated IR.}
\KwOut{Best allocation plan, $\mathcal{A}_{best}$.}

$\mathcal{H}_{units} \leftarrow$ all state element groups $(s, f)$ from NFs in $\mathcal{N}_{all}$\;
Sort $\mathcal{H}_{units}$ descending by key $k = (R_{s,f} \cdot |\mathcal{I}_s|) / size(s,f)$\;

Initialize $\mathcal{A}_{candidate}$ with all $(s,f) \in \mathcal{H}_{units}$ mapped to $DRAM$\;
$\mathcal{A}_{best} \leftarrow \{\mathcal{A}_{candidate}, Predictor(\mathcal{A}_{candidate})\}$\;
\ForEach{$(s, f, R_{s,f}, \mathcal{I}_s)$ in $\mathcal{H}_{units}$}{
  $\mathcal{A}_{try} \leftarrow \mathcal{A}_{candidate}$ with $\mathcal{A}_{try}(s, f, R_{s,f}, \mathcal{I}_s) = SRAM$\;

  \If{\textbf{not} $\textit{ResourceFail}(\mathcal{A}_{try})$}{
    $P = \textit{Predictor}(\mathcal{A}_{try})$\;
    $\mathcal{A}_{candidate} \leftarrow \{\mathcal{A}_{try}, P\}$\;
    \If{$\textit{Objective}(\mathcal{A}_{candidate})$ > $\textit{Objective}(\mathcal{A}_{best})$}{
      $\mathcal{A}_{best} \leftarrow \mathcal{A}_{candidate}$\;
    }
  }
}

\Return $\mathcal{A}_{best}$\;
\end{footnotesize}
\end{algorithm}

\begin{table*}[t]
\centering
\footnotesize
\renewcommand{\arraystretch}{1.15}
\begin{tabular}{@{}lllll@{}}
\toprule
\revised{State / access path} & \revised{Memory tier} & \revised{Primitives per visit} & \revised{Relative rate} & \revised{Atomicity} \\
\midrule
\revised{\textit{ConnTable}: index lookup} & \revised{IMEM (SRAM)} & \revised{One 4-byte atomic read} & \revised{1.0} & \revised{Memory engine} \\
\revised{\textit{ConnTable}: cached data} & \revised{IMEM (SRAM)} & \revised{One 32-byte atomic read} & \revised{0.8} & \revised{Memory engine} \\
\revised{\textit{ConnTable}: uncached data} & \revised{EMEM (DRAM)} & \revised{One 32-byte atomic read} & \revised{0.2} & \revised{Memory engine} \\
\revised{\textit{DipCntr}} & \revised{IMEM (SRAM)} & \revised{One 4-byte atomic increment} & \revised{1.0} & \revised{Memory engine} \\
\revised{\{\textit{MinLoad}, \textit{MinDip}\}} & \revised{IMEM (SRAM)} & \revised{Two 4-byte bulk reads and two writes} & \revised{1.0} & \revised{Software spinlock} \\
\bottomrule
\end{tabular}
\caption{\revised{NATLB memory-operation mix for an illustrative candidate layout on Agilio: the hottest quarter of \textit{ConnTable} is cached in SRAM, while both counter blocks reside entirely in SRAM. Rates follow the running example's access profile, not the fraction of entries placed in each tier. Full primitive parameters appear in Table~\ref{tab:state_parameters}.}}
\label{tab:natlb_candidate}
\end{table*}

\para{Optimized Allocation via Ordering.}
Finding the optimal placement of all state element groups is an exponential problem with a search space of roughly $O(2^{\sum_{s} |\mathcal{T}_s|})$, where $\sum|\mathcal{T}_s|$ is the total number of state element groups. 
To make this tractable, \app employs a \textit{Global Ordering} heuristic (Algorithm~\ref{alg:allocation}). 
Instead of a brute-force search, the algorithm performs a single constructive pass.
First, it establishes a baseline plan by mapping all state element groups to DRAM (Line 3). 
It then sorts all groups in descending order of their \revtag{R16}\revised{\textit{memory-access density}} with key $k = (R_{s,f} \cdot |\mathcal{I}_s|) / size(s,f)$,
\revised{prioritizing groups with more profile-weighted memory accesses per byte of stored state (Line 2).
This directs limited SRAM capacity toward frequently accessed state, while the Predictor evaluates the resulting throughput.}
The algorithm then greedily iterates through this sorted list, attempting to promote each group to the high-performance SRAM (Line 6). 
A promotion is committed if it does not violate SRAM's capacity constraints (Line 7). 
At each step, the algorithm tracks the best-performing plan found so far by consulting the \textbf{Predictor} (Lines 8-11). 
This heuristic-driven search reduces the complexity from exponential to linearithmic, $O((\sum_{s} |\mathcal{T}_s|)\cdot \log({\sum_{s} |\mathcal{T}_s|}))$.

\para{\revised{Evaluating a Candidate.}}
\revised{At\revtag{R2} each Predictor call in Algorithm~\ref{alg:allocation}, the candidate layout determines the memory operations and their relative rates, as illustrated in Table~\ref{tab:natlb_candidate}.
Using the target's profiled performance curves, the Predictor estimates the NF's aggregate throughput ceiling (\S\ref{sec-performance}).
It also estimates each shared lock instance's optimal thread count and sustainable access rate, flagging instances whose sampled demand exceeds this rate (\S\ref{subsec:perform:lock}).
For NATLB, placing the \{\textit{MinLoad}, \textit{MinDip}\} pair affects how long each thread holds the lock, while placing \textit{ConnTable} affects the time between lock acquisitions; both therefore influence this thread count.
Appendix~\ref{subsec:prediction_example} details the calculation.}

\noindent\textbf{Auxiliary IR Generation.}
Based on the final plan $\mathcal{A}$, the Synthesizer generates auxiliary IR that instructs the downstream compiler backend on how to implement the plan:
First, it inserts spinlock primitives for state blocks identified by the Analyzer as requiring them.
Second, it generates steering configuration specifying the optimal number of threads (i.e., $N_{critical}$ returned by Predictor, detailed in \S\ref{subsec:perform:lock}) for servicing each locked state, 
which can be enforced by hardware steering rules or software dispatchers (Appendix~\ref{subsec:backend:steer}).
Third, for partitioned states, it generates the necessary software cache logic, including the index remap table and cache management procedures. 
This auxiliary IR, combined with the original NF logic, provides a complete specification for the open-source compiler's backend to generate the final NIC source code, 
which is then compiled into firmware by the vendor toolchain.

%% file: 6_1-Performance_Model.tex
\section{Performance Model}
\label{sec-performance}

Within \app's compilation workflow, the performance model quantitatively evaluates allocation plans proposed by the Synthesizer (\S\ref{subsec:synthesizer}). 
\revtag{R3}\revised{With memory layouts fixed at compile time by the target hardware, the model guides state allocation for sustained processing capacity by predicting steady-state throughput under the sampled traffic profile.
Predicting transient effects from cache warm-up, individual evictions, and short-lived traffic bursts offers limited benefit; they are instead handled at runtime (\S\ref{sec:runtime}).}
This section details our core approach, first covering our core methodology (\S\ref{subsec:perform:methodology}) and then the lock-free (\S\ref{subsec:perform:roofline}) and lock contention (\S\ref{subsec:perform:lock}) models.

\vspace{-1.0\baselineskip}
\subsection{Modeling Methodology}
\label{subsec:perform:methodology}
On multi-threaded SoC SmartNICs like BlueField 3 and Agilio, performance is dictated by the interaction between parallel worker threads and the shared memory subsystem. 
As our profiling confirms, global shared memory accesses dominate processing overhead and frequently become the system bottleneck (Table~\ref{tab:agilio_memory}). 
Predicting the peak throughput of NFs therefore requires accurately modeling when and how the memory subsystem throttles overall performance.

Our approach is based on a divide-and-conquer strategy. 
Instead of modeling an entire complex NF monolithically, 
we first decompose any NF into a mix of primitive, hardware-level memory operations (e.g., an atomic read from on-chip SRAM, a bulk write to off-chip DRAM).
Then, we characterize the performance of each primitive operation in isolation via one-time, offline microbenchmarking, creating a database of performance curves for each target NIC.
Last, we compose the performance of these primitives based on the operation intensity dictated by the NF logic, the mix pattern derived from the workload trace, 
and the performance curves selected by a given allocation plan, to predict the final throughput.

The core of our model is to analyze this as a supply-and-demand problem: 
worker threads demand memory access by executing instructions ($\mathcal{I}_s$) from state blocks, while the NIC's memory engines supply this access. 
A performance bottleneck emerges when the aggregate demand from parallel threads exceeds an engine's service capacity (its supply), causing it to saturate and throttle the entire system.
Inspired by SoC performance modeling works like Gables~\cite{Gables}, 
we recognize that because memory engines (e.g., for on-chip vs. off-chip memory) operate in parallel, 
we need to analyze this supply-demand balance for each engine independently to find the most constrained one. 
Another key factor in this model is software locking. 
Required for state consistency, locks serialize execution, fundamentally altering the parallelism assumption. 
This insight motivates decomposing our prediction problem into two questions:

\para{Q1:} Under ideal, lock-free conditions, what is the maximum aggregate packet throughput a SmartNIC can sustain for a given NF's workload?

\para{Q2:} For a specific lock-protected state block, what is the maximum access rate the SmartNIC can support before lock contention saturates its processing capability?

Answering Q1 provides the global throughput ceiling for capacity planning, while answering Q2 prevents localized contention bottlenecks. 
We detail our models for each.

\subsection{Modeling Lock-Free Throughput}
\label{subsec:perform:roofline}

{
\begin{figure}[!tp]
\centering
\setlength{\belowcaptionskip}{-10pt}
\minipage{0.48\textwidth}
    \begin{subfigure}{0.48\textwidth}
        \includegraphics[width=\linewidth]{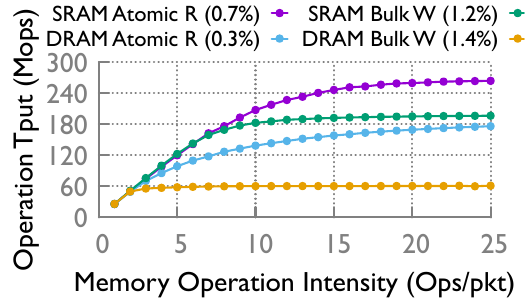}
        \vspace{-15pt}
        \caption{Agilio.}
    \end{subfigure}
    \hspace*{\fill}
    \begin{subfigure}{0.48\textwidth}
        \includegraphics[width=\linewidth]{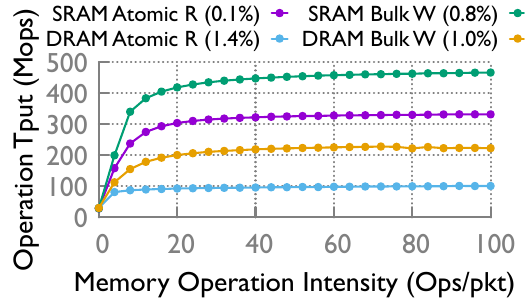}
        \vspace{-15pt}
        \caption{BlueField 3.}
    \end{subfigure}
\endminipage
\caption{SoC SmartNICs' example performance curves.}
\label{fig:mem_roofline}
\vspace{-6pt}
\end{figure}
}

To answer Q1, we build a database of primitive performance curves and provide a method for composing them.

\para{Building the Primitive Curve Database.} 
To quantify the supply side of our model, we characterize the relationship between the demand placed on a memory engine and the packet throughput it can sustain. 
We adapt the Roofline model concept~\cite{WilliamsWP09} to build performance curves, capturing this relationship for each primitive memory operation.
We introduce the concept of \textit{memory operation intensity} ($i$), the number of times a specific memory operation is executed per packet. 
As shown in Figure~\ref{fig:mem_roofline}, we empirically measure the total memory operation throughput $M$ as we vary the intensity ($i$). 
The resulting curves consistently show three phases: an initial linear growth where memory latency is hidden by thread parallelism, 
a transitional phase of diminishing returns as latency begins to outpace computation, 
and a final saturation "roof" where performance is bound by a hardware limit-either a memory engine's command rate or the underlying memory bandwidth.

The shape of a performance curve is highly specific to the executed operation, as different primitives exhibit unique latency and parallelism characteristics. 
Therefore, we generate these curves by running microbenchmarks for each primitive operation type, 
defined by parameters including memory tier $t$ (SRAM/DRAM), operation $o$ (read/write), memory engine $e$ (atomic/bulk), data transfer size $s$, 
and memory array size $n$ (relevant for caching or banking effects). 
Different combinations of ($t, o, e, s, n$) yield distinct performance profiles.
To manage the large parameter space, we profile a representative set of common values.
For data size $s$, we test the discrete sizes supported by the memory interface (e.g., 4/16/64/128 bytes). 
For array size $n$, we profile sizes spanning megabytes to gigabytes and interpolate for intermediate values. 
The number of profiled primitives is determined by the NIC's exposed memory interfaces (320 on Agilio, 288 on BlueField 3); 
hardware offering more granular access control typically produces more profiles than hardware with a unified model.

\para{Analytical Model.} 
We fit our measured data to a physically-motivated model derived from performance fundamentals.
For a given primitive operation, let $\Delta t_m$ be its physical memory latency and $\Delta t_0$ be the base per-packet processing overhead (e.g., hardware-accelerated packet parsing, framework costs).
The primitive's maximum operation throughput, $M_{roof}$, defines the engine's Memory-Level Parallelism (MLP) via Little's Law~\cite{little2008little}: 
$MLP \approx M_{roof} \times \Delta t_m$, representing the maximum number of in-flight requests in executing this primitive.
The packet throughput for a workload with intensity $i$ is then the engine's concurrency (MLP) divided by the per-packet service time ($\Delta t_{task}(i)=\Delta t_0+\Delta t_m*i$).
We convert this packet throughput ($T$) to memory operation throughput ($M(i)$) by multiplying by $i$, as each packet involves $i$ operations:
\begin{equation}
\footnotesize
\begin{split}
M_{t, o, e, s, n}(i) = T \times i = \frac{MLP}{\Delta t_{task}(i)} \times i = \frac{M_{roof} \cdot \Delta t_m \cdot i}{\Delta t_0 + \Delta t_m \cdot i}
\end{split}
\end{equation}
To account for latency hiding via thread switching at low intensities, we model an effective memory latency that approaches zero when $i$ is low 
and approaches the physical latency $\Delta t_m$ when $i$ is high. 
We use an exponential penalty term $e^{-b/i}$ for this, yielding our final fitted model:
\begin{equation}
\footnotesize
    \label{eq:mem_roofline}
    \begin{split}
    M_{t, o, e, s, n}(i) & = \frac{M_{roof} \cdot \Delta t_m \cdot i}{\Delta t_0 + \Delta t_m \cdot e^{-\frac{b}{i}} \cdot i}
                        = \frac{M_{roof} \cdot i}{a+ e^{-\frac{b}{i}} \cdot i},
    \end{split}
\end{equation}
where $a = \Delta t_0/\Delta t_m$. The parameters $M_{roof}$, $a$, and $b$ are fitted from microbenchmark data for each primitive. 
The parameter $b$ implicitly captures caching effects: a higher $b$ indicates better data locality for the array size $n$ on caching architectures like BlueField 3. 
This model fits the data well, with average errors of $0.77\%\pm 1.17\%$ for Agilio and $1.67\%\pm 2.11\%$ for BlueField 3 (example errors in Figure~\ref{fig:mem_roofline} legends).

\para{Composing Primitives for NFs.} 
To predict the performance of a complete NF or multiple NFs, the Predictor composes the performance curves of its constituent primitive operations. 
Given an allocation plan, the Synthesizer provides the Predictor with a mix of memory operations and their execution rates, grouped by the hardware engine they share. 
For each engine, the Predictor derives a composite performance curve by first calculating a weighted-average memory latency ($\Delta t_{mix}$) for that specific mix. 
This latency informs a new analytical curve generated via a synthesize-and-fit process. 
The Predictor then identifies the engine with the lowest composite performance as the system-wide bottleneck, 
from which the final lock-free packet throughput is calculated. 
The detailed methodology is presented in Appendix~\ref{subsec:perform:hybrid}.
The final result is the predicted throughput for the plan under ideal, parallel conditions, which we denote as $T_{lockfree}$, providing the answer to Q1.

\vspace{-0.5\baselineskip}
\subsection{Modeling Lock Contention Impact}
\label{subsec:perform:lock}

{
\begin{figure}[!tp]
\vspace{-3pt}
\setlength{\abovecaptionskip}{-0.4pt}
\centering
\begin{subfigure}{0.49\linewidth}
    \includegraphics[width=\linewidth]{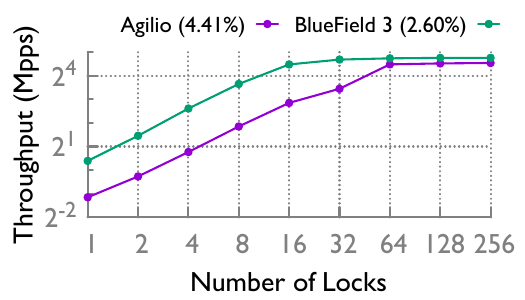}
    \vspace{-15pt}
    \caption{Independent lock scaling.}
    \label{fig:lock_scaling}
\end{subfigure}
\hspace*{\fill}
\begin{subfigure}{0.49\linewidth}
    \centering
    \includegraphics[width=0.9\linewidth]{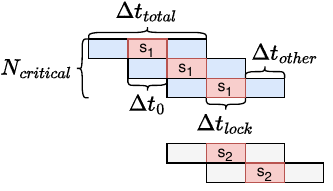}
    \caption{Shared-lock contention.}
    \label{fig:lock_diagram}
\end{subfigure}
\caption{Model for lock scaling and lock contention.}
\label{fig:lock}
\vspace{-15pt}
\end{figure}
}

To answer Q2 (the maximum throughput for a locked state), we model how the serialization caused by a single, contended software lock impacts the overall system performance.

Before analyzing the single-lock case, we first establish the overall performance ceiling with locks.  
The lock scaling experiment in Figure~\ref{fig:lock_scaling} demonstrates that when a workload is distributed across many independent, 
non-interfering lock instances, the aggregate throughput scales nearly linearly until it saturates. 
This saturation point empirically validates that the system's maximum capacity is bounded by its ideal, contention-free performance, $T_{lockfree}$ (Q1's answer), 
thereby establishing $T_{lockfree} = N_{total}/\Delta t_{total}$ as the theoretical supply of the entire system, where $ N_{total}$ represents the total number of available, independent processing threads.

In contrast, performance collapses when multiple threads compete for the same lock, as illustrated in Figure~\ref{fig:lock_diagram}. 
Because the critical section can only service one thread at a time, other threads arriving at the lock must queue and wait. 
The key to modeling this is to quantify how many parallel threads can be kept busy by the non-critical section work while one thread occupies the lock. 
We define this as $N_{critical}$, the maximum number of threads that can effectively utilize a single lock instance without significant serialization. 
It is determined by the ratio of total processing time to the critical section time: $N_{critical} \approx \Delta t_{total}/\Delta t_{lock}$.
\revtag{R16}\revised{In its current form, the model does not explicitly account for lock convoy dynamics, as it uses execution costs to estimate steady-state parallelism for compile-time layout exploration.
Such dynamics may introduce additional waiting even with ample work outside the critical section and make $N_{critical}$ optimistic.}

With these concepts, we can derive the maximum sustainable throughput for a single locked state, $T_{lock\_max}$, 
by relating its fundamental limit ($1/\Delta t_{lock}$) to the overall system capacity. By substituting the terms defined above, 
we can express $T_{lock\_max}$ as a fraction of the total system throughput:
\begin{equation}
\footnotesize
\label{eq:lock_max_throughput}
T_{lock\_max} = \frac{1}{\Delta t_{lock}} = \frac{N_{total}}{\Delta t_{total}} \cdot \frac{\Delta t_{total}}{\Delta t_{lock}} \cdot \frac{1}{N_{total}} = T_{lockfree} \times \frac{N_{critical}}{N_{total}}
\end{equation}

We estimate $N_{critical}$ using parameters derived from our analytical models. 
As detailed in Appendix~\ref{subsec:perform:hybrid}, we can derive the parameters $a_{total}$ and $a_{lock}$, 
which represent the ratio of program-independent processing overhead ($\Delta t_0$) to program-dependent memory service time.
Specifically, $a_{total}$ is the ratio of this overhead to the total memory service time for the full packet path ($a_{total}=\Delta t_0/(\Delta t_{lock}+\Delta t_{other})$), 
while $a_{lock}$ is the ratio of overhead to the service time of just the critical section ($a_{lock}=\Delta t_0/\Delta t_{lock}$). 
This allows us to calculate $N_{critical}$ as:
\begin{equation}
\footnotesize
\label{eq:ncritical}
N_{critical} = a_{lock} \left(1 + \frac{1}{a_{total}}\right)
\end{equation}

The results of contention model provide two crucial outputs. 
First, the predicted $N_{critical}$ value is used to constrain the assigned parallelism (i.e., thread group size) for the locked state, 
preventing wasted cycles from excessive contention. 
Second, the calculated $T_{lock\_max}$ is surfaced to the operator as a key diagnostic,  
allowing her to determine if the expected workload 
(based on their target throughput and the state's relative access rate, $R_{s,f}$) 
will saturate the lock. If so, Vela's prediction serves as a critical bottleneck warning.

%% file: 6-Runtime.tex
\vspace{-10pt}
\section{\app Runtime}
\label{sec:runtime}

{
\setlength{\abovecaptionskip}{2pt}
\setlength{\belowcaptionskip}{-2pt}
\begin{figure*}[tp]
  \centering
  \minipage{0.40\textwidth}
    \centering
    \includegraphics[width=\linewidth]{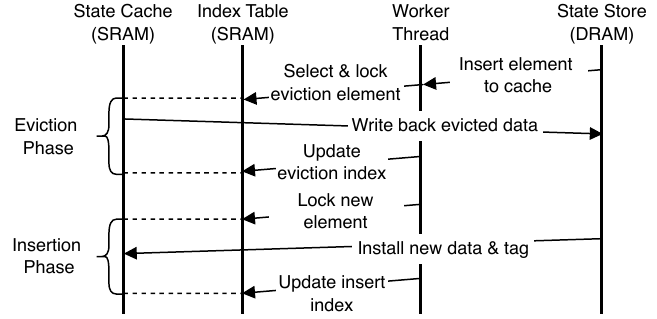}
    \vspace{-6pt}
    \caption{\app's state cache updating protocol.}
    \label{fig:migration}
  \endminipage
  \hspace*{\fill}
  \minipage{0.58\textwidth}
    \centering
    \begin{subfigure}{0.49\textwidth}
      \includegraphics[width=0.95\linewidth]{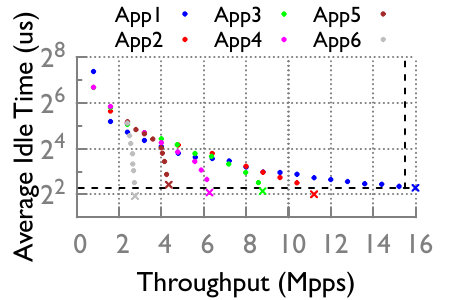}
      \vspace{-2pt}
      \caption{Agilio.}
    \end{subfigure}
    \hspace*{\fill}
    \begin{subfigure}{0.49\textwidth}
      \includegraphics[width=0.95\linewidth]{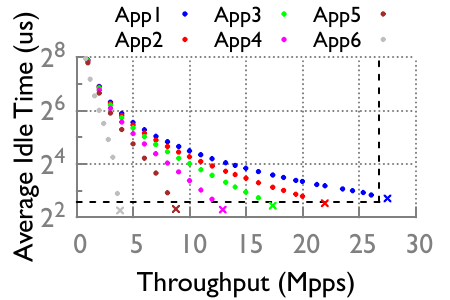}
      \vspace{-2pt}
      \caption{BlueField 3.}
    \end{subfigure}
    \caption{The idle time distribution for different applications.}
    \label{fig:idle_time}
  \endminipage
  \vspace{-12pt}
\end{figure*}
}

\app's compile-time framework (\S\ref{sec:backend}) produces a static allocation plan optimized for a given traffic profile. 
\revtag{R4}\revised{The runtime's primary role is to manage the software cache for compiler-partitioned state, adapting its contents to live traffic without changing the compiled layout (\S\ref{subsec:runtime:caching}).
Load monitoring and mitigation provide a supporting safety net when traffic exceeds the deployed plan's processing capacity (\S\ref{subsec:runtime:monitoring}).}

\vspace{-0.5\baselineskip}
\subsection{Runtime State Caching}
\label{subsec:runtime:caching}
For large state objects that the compiler partitions across memory tiers (e.g., \textit{ConnTable}), 
\revtag{R3}\revised{the runtime dynamically populates the allocated SRAM cache with hot state elements without changing the compiled memory layout.
Under a stable access profile, this moves the cache toward the steady-state access mix assumed by the Predictor.
Warm-up and cache swaps can temporarily reduce throughput, while rap\-idly shifting hot sets may prevent the cache from reaching this regime, as discussed below.}

\para{Cache Access Mechanism.}
The software cache logic is explicit and operates on whole state objects. 
For a partitioned state, the compiler generates two arrays: a smaller State Cache in SRAM and a larger State Store in DRAM that holds the complete object data. 
To mediate access, it also generates an Index Table in SRAM, mapping each element's logical index to a physical index into one of the two arrays.
Initially, all entries point to their home locations in the State Store, leaving the State Cache ready to be populated at runtime.
To read a state element, a worker thread checks the corresponding Index Table entry's lock flag and waits if necessary before reading the mapping without acquiring the lock.
To modify an element, the thread acquires the entry's lock before reading the mapping and holds it until the write completes.
The physical index directs the access to the State Cache if it falls within the cache's bounds, or to the State Store otherwise.

\para{Hot Entry Detection.} 
To identify promotion candidates from DRAM, each processing core maintains a shared, lightweight sketch~\cite{HeavyKeeper} 
in memory accessible to its threads (e.g., Agilio's CTM or BlueField 3's DPA private memory) to efficiently track the access frequency of DRAM-resident state elements. 
This sketch is cleared periodically, and each thread uses it to identify its own "hottest" candidate for promotion.

While sketches excel at finding heavy hitters, there is no equivalent low-cost method for identifying a cold element in SRAM to evict. 
To circumvent this, we use the running average of the access counts of all historically promoted entries as a dynamic admission threshold. 
A new DRAM candidate is admitted only if its access count exceeds this average, ensuring new entrants are at least as active as their predecessors were at the time of admission.
However, this historical average can become a stale barrier. 
If traffic shifts to a new, less frequent set of hot flows, this high threshold could prevent them from ever entering the cache. 
We add a simple heuristic to break this deadlock: after three consecutive failed promotion attempts, a thread forcibly promotes its hottest DRAM candidate. 
\revtag{R12}\revised{Under rapidly shifting hot sets, however, newly promoted entries may become cold before enough cache hits accrue to offset the lookup and cache-swap overheads.}

\para{Cache Update Protocol.} 
When a thread identifies a hot element, it initiates a cache update, 
as illustrated in Figure~\ref{fig:migration}, to swap it with an element currently in the State Cache. 

\squishlist
\item\textbf{Eviction Phase:} The worker thread first selects a victim entry for eviction from the State Cache using a simple random selection policy.
It then acquires an exclusive lock on the victim's corresponding entry in the SRAM-based Index Table. 
This lock forces any other threads that subsequently look up this index to block. With the lock held, 
the worker \revised{writes back the victim's data, including updates accumulated in SRAM,} to its home location in the State Store (DRAM),
updates the Index Table entry to point to this DRAM location, and finally releases the lock.
Blocked threads can now proceed and will be correctly directed to the data in DRAM.

\item\textbf{Insertion Phase:} The worker thread then acquires a lock on the Index Table entry for the new hot element (which currently points to DRAM). 
It atomically installs the new element's data and logical-index tag in the now-vacant slot in the State Cache.
Finally, it updates the Index Table to point to this new SRAM location and releases the lock.
\squishend

\revtag{R13}\revised{A reader may obtain an SRAM mapping just before eviction locks the corresponding Index Table entry and use it after the cache slot has been reassigned.
To detect such stale accesses, each State Cache slot stores its element's logical index alongside its data.
The reader retrieves both in a single atomic read and accepts the data only if the tag matches the requested logical index; otherwise, it rereads the Index Table and retries.
Writes hold the corresponding Index Table entry's lock throughout the access, preventing eviction or insertion from moving the element before the write completes.}

\vspace{-0.5\baselineskip}
\subsection{Load Monitoring and Mitigation}

\label{subsec:runtime:monitoring}
While state caching optimizes performance in normal cases, 
the runtime also needs a safety mechanism to prevent packet loss when the NIC becomes computationally overloaded due to live traffic mismatching the static plan's assumptions.

\para{Overload Indicator.} To trigger reactive measures, the runtime needs a reliable, low-overhead overload indicator. 
While prior approaches often monitor packet processing latency, establishing effective thresholds can be challenging, 
as latency varies significantly with application logic and traffic patterns~\cite{ipipe, DBLP:conf/sosp/QiuXHKLNC21}. 
\app instead leverages run-to-completion processing on worker threads,
which makes the per-thread idle time a robust, program-agnostic indicator. 
For each thread, a timestamp is recorded after it completes processing a packet ($t_{end}$) and before it begins the next ($t_{start}$). 
The idle time, $t_{idle} = t_{start} - t_{end}$, is a direct measure of available processing cycles. 
The host-side Monitor periodically reads these idle time values from shared NIC memory, 
and if the calculated average consistently falls below a pre-determined, NIC-specific threshold $T$, the system is considered overloaded. 
Figure~\ref{fig:idle_time} validates this approach, showing a clear, application-independent separation between non-overloaded and overloaded (marked by packet drops) conditions.
The dotted lines define a "safe operating zone" bounded by the overload threshold $T$ and the NIC's maximum throughput.

\para{Reactive Overload Mitigation.} 
When the Monitor detects an overload event (average $t_{idle}$ < $T$), it signals the host-side Orchestrator. 
\revtag{R9}\revised{The Orchestrator reads the on-NIC sketches to identify large flows and uses the application's IR to map them to state element groups $(s,f)$ (\S\ref{subsec:synthesizer}).
For each selected high-load group, it temporarily redirects all traffic accessing that group from the SmartNIC to available host/Arm cores.
In the running NATLB example, this includes all flows accessing the pair \{\textit{MinLoad}[$f$], \textit{MinDip}[$f$]\} for the selected server-group index $f=\textit{vip\_id}$.
Each redirected group is thus served exclusively by the SmartNIC or the host/Arm, avoiding cross-device concurrent access and locking.}
This acts as a safety valve, reducing the NIC's workload until its average idle time recovers to the safe zone. 
For persistent traffic shifts, a full recompilation is triggered (Figure~\ref{fig-arch}).

%% file: 7-Eval.tex
\vspace{-0.5\baselineskip}
\section{Evaluation}
\label{sec:eval}

\begin{figure*}[tp]
\centering
\setlength{\belowcaptionskip}{-10pt}
\minipage{0.95\textwidth}
\centering
    \begin{subfigure}{0.48\textwidth}
    \includegraphics[width=\linewidth]{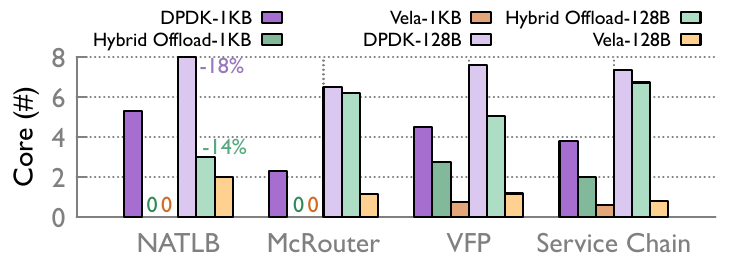}
    \end{subfigure}
    \begin{subfigure}{0.48\textwidth}
    \includegraphics[width=\linewidth]{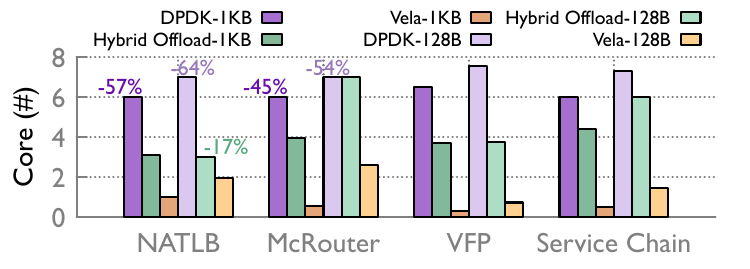}
    \end{subfigure}
\endminipage
\vspace{-12pt}
\caption{Core usage for Agilio (left) and BlueField 3 (right) SmartNICs across different network applications. 
1KB packets target the SmartNIC's line rate, while 128B packets target 15~Mpps on Agilio and 30~Mpps on BlueField 3.}
\label{fig:e2e}
\end{figure*}

Our evaluation investigates \app's: 
(1) end-to-end application performance (\S\ref{subsec:eval:e2e}); 
(2) compile-time allocation plan adaptability across diverse traffic patterns (\S\ref{subsec:eval:compiler}); 
and (3) runtime adjustment capability for network dynamics (\S\ref{subsec:eval:runtime}).

\noindent\textbf{Testbeds.} 
We evaluated \app on two SoC SmartNICs: a 40~Gbps Netronome Agilio CX SmartNIC and a 200~Gbps NVIDIA BlueField 3 DPU. 
On Agilio testbed, excess traffic is handled by the host CPU (2x24-core CPU, 128~GB RAM). 
On BlueField 3 testbed, we utilize its integrated Arm subsystem (16 cores, 32~GB RAM) for excess traffic, 
a common DPU use case that aims to minimize host CPU consumption. 
Programs were developed with vendor SDKs (Netronome NFP SDK 6.1.0, NVIDIA DOCA FlexIO 24.10) 
and default to using the maximum available hardware threads (216 FPC threads on Agilio, 254 DPA threads on BlueField 3).

\noindent\textbf{Implementation and Workloads.}
We integrated \app with p4c~\cite{p4c}. 
Its output memory allocation and steering configuration guides the generation of 
MicroC~\cite{micro_c} for Agilio and DPA C~\cite{flexio} for BlueField 3.
We used the CIC-IDS2017 network trace~\cite{cic-ids2017}, 
containing 39~M packets across 1.3~M flows (distribution in Figure~\ref{fig:fb}), 
which we replayed to measure the application throughput. 
For McRouter, we generated Memcached traffic using YCSB~\cite{Cooper2010YCSB} with a
\revtag{R16}\revised{Workload A read/update mix (50\% each)},
128-byte keys, and a Zipf (0.99 skew) request pattern, following~\cite{lim2014mica}.

{
\begin{figure*}[tp]
\setlength{\belowcaptionskip}{-8pt}
\centering
\minipage{0.95\textwidth}
\centering
  \begin{subfigure}{0.24\textwidth}
    \includegraphics[width=\linewidth]{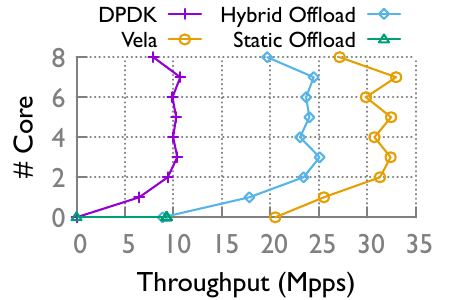}
    \vspace{-18pt}
    \caption{Core vs throughput.}
    \label{fig:natlb:core_thput:bf3}
  \end{subfigure}%
  \begin{subfigure}{0.24\textwidth}
    \includegraphics[width=\linewidth]{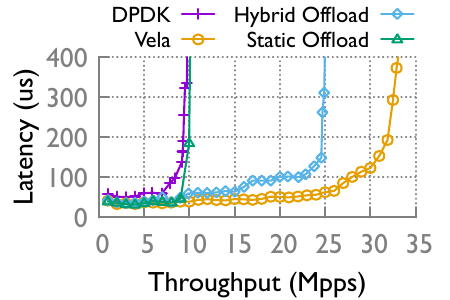}
    \vspace{-18pt}
    \caption{Latency vs throughput.}
    \label{fig:natlb:lat_thput:bf3}
  \end{subfigure}%
  \begin{subfigure}{0.24\textwidth}
    \includegraphics[width=\linewidth]{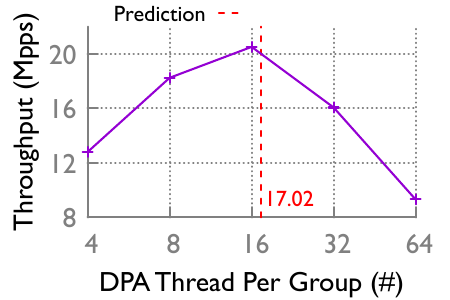}
    \vspace{-18pt}
    \caption{Impact of thread contention.} 
    \label{fig:natlb:thread_thput:bf3}
  \end{subfigure}
  \begin{subfigure}{0.24\textwidth}
    \includegraphics[width=\linewidth]{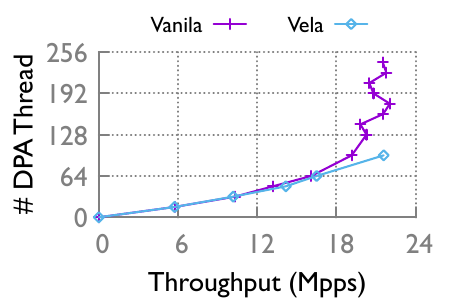}
    \vspace{-18pt}
    \caption{Impact of thread workload.} 
    \label{fig:natlb:thread_table:bf3}
  \end{subfigure}
  \vspace{-2pt}
\caption{Accelerating NATLB on BlueField 3 and thread usage tuning based on the performance model.}
\label{fig:natlb:bf3}
\endminipage
\vspace{-4pt}
\end{figure*}
}

{
\begin{figure}[!tp]
\vspace{-3pt}
\setlength{\abovecaptionskip}{-0.4pt}
\setlength{\belowcaptionskip}{-1pt}

\minipage{0.48\textwidth}
\centering
    \begin{subfigure}{0.49\textwidth}
        \includegraphics[width=\linewidth]{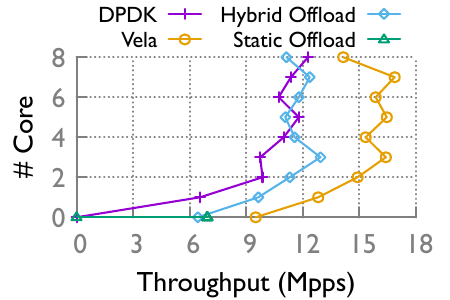}
        \vspace{-16pt}
        \caption{Core vs throughput.}
        \label{fig:natlb:core_thput:agilio}
    \end{subfigure}
    \begin{subfigure}{0.49\textwidth}
        \includegraphics[width=\linewidth]{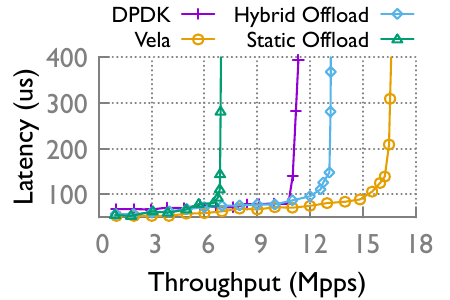}
        \vspace{-16pt}
        \caption{Latency vs throughput.}
        \label{fig:natlb:lat_thput:agilio}
    \end{subfigure}
    \caption{Accelerating NATLB on Agilio.}
    \label{fig:natlb:agilio}
\endminipage
\end{figure}
}

{
\begin{figure}[!tp]
\vspace{-3pt}
\setlength{\abovecaptionskip}{-0.4pt}
\setlength{\belowcaptionskip}{-12pt}

\minipage{0.48\textwidth}
\centering
  \begin{subfigure}{0.49\textwidth}
      \includegraphics[width=\linewidth]{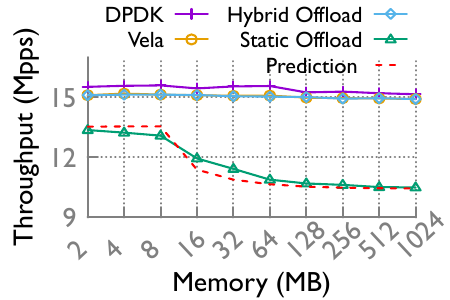}
      \vspace{-18pt}
      \caption{Agilio.}
      \label{fig:mcrouter_thput:agilio}
  \end{subfigure}
  \hspace*{\fill}
  \begin{subfigure}{0.49\textwidth}
      \includegraphics[width=\linewidth]{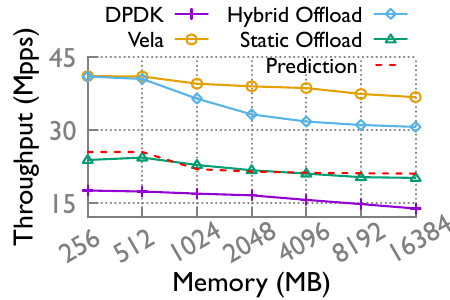}
      \vspace{-18pt}
      \caption{BlueField 3.}
      \label{fig:mcrouter_thput:bf3}
  \end{subfigure}
  \hspace*{\fill}
  \caption{Accelerating McRouter on Agilio and BlueField 3.}
  \label{fig:mcrouter}
\endminipage
\vspace{-5pt}
\end{figure}
}

{
\begin{figure*}[tp]
\setlength{\abovecaptionskip}{8pt}
\setlength{\belowcaptionskip}{-6pt}
\centering
\minipage{0.95\textwidth}
\centering
  \begin{subfigure}{0.24\textwidth}
    \includegraphics[width=\linewidth]{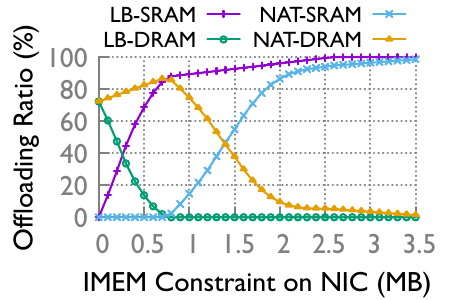}
    \vspace{-18pt}
    \caption{Size of fast memory.}
    \label{fig:sensitivity-agilio:mem_size}
  \end{subfigure}%
  \begin{subfigure}{0.24\textwidth}
    \includegraphics[width=\linewidth]{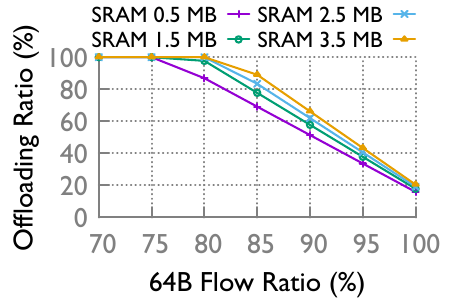}
    \vspace{-18pt}
    \caption{Traffic load.}
    \label{fig:sensitivity-agilio:traffic}
  \end{subfigure}%
  \begin{subfigure}{0.24\textwidth}
    \includegraphics[width=\linewidth]{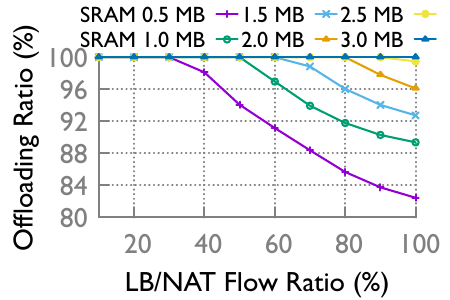}
    \vspace{-18pt}
    \caption{Intra-NF traffic distribution.} 
    \label{fig:sensitivity-agilio:state}
  \end{subfigure}
  \begin{subfigure}{0.24\textwidth}
    \includegraphics[width=\linewidth]{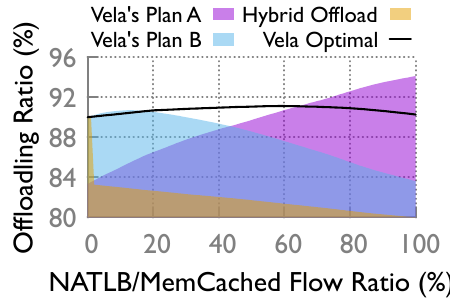}
    \vspace{-18pt}
    \caption{Inter-NF traffic distribution.} 
    \label{fig:sensitivity-agilio:ratio}
  \end{subfigure}
  \vspace{-2pt}
\caption{Offloading ratio under different factors on the Agilio SmartNIC.} \label{fig:sensitivity-agilio}
\endminipage
\vspace{-10pt}
\end{figure*}
}

\subsection{Application Performance}
\label{subsec:eval:e2e}
We accelerated the following applications with \app:

\noindent\textbf{NATLB}~\cite{pismenny2022nicmem} implements the functionality described in our Motivating Example (Figure~\ref{fig:motiv_example}), 
scaled to simulate 4096 backend servers (organized into 256 groups of 16), maintaining per-group counters to track the minimum load.

\noindent\textbf{McRouter}~\cite{mcrouter} routes Memcached requests using an on-NIC hash table. 
To optimize lookups, it leverages pre-computed key hashes included in packet headers, 
a technique inspired by high-performance key-value systems~\cite{lim2014mica,ghigoff2021bmc}. 
The hash identifies a bucket, where a subsequent key comparison finds the mapping to the correct backend server. 

\noindent\textbf{VFP}~\cite{firestone2017vfp} implements flexible VM network policies. 
Our implementation applies a sequence of NFs including traffic monitoring, firewalling, and load balancing.

\noindent\textbf{Service Chain} inspects Memcached requests on a ToR switch, 
sending them to a Count-min sketch for telemetry while a parallel branch applies NATLB to other general traffic.

\para{Baselines.} 
We compare \app against two intuitive yet suboptimal baselines, in addition to a software-only DPDK implementation that establishes a CPU-only performance anchor.
(1) \textbf{Static Offload} models a naive, traffic-aware offloading strategy by statically offloading all NF states 
and greedily placing states with the highest \textit{memory-access density} (\S\ref{subsec:synthesizer}) into SRAM until full.
Large states are never partitioned; if a state object exceeds available SRAM, it is placed entirely in DRAM. 
This baseline operates under the optimistic assumption that all offloaded operations can run at line rate 
and is designed to show how ignoring hardware performance models leads to packet drops.
(2) \textbf{Hybrid Offload} models a common operational workaround by adding a runtime safety net to the Static Offload plan. 
It monitors the NIC's overall load and, upon detecting overload, dynamically steers entire traffic flows back to the CPU/Arm for processing (\S\ref{subsec:runtime:monitoring}). 
This baseline is designed to quantify the high CPU cost required to compensate for a statically suboptimal offload plan.

\para{Experimental Setup and Results.}
We evaluate a bandwidth-intensive test using 1KB packets 
and a processing-intensive test using 128B (targets in Figure~\ref{fig:e2e} caption). 
We use the number of cores required to meet performance targets without packet drops as the main metric.
We use BlueField 3's Arm cores, while on Agilio, we utilize up to 8 NUMA-local CPU cores. 
For the Service Chain, traffic is distributed evenly between branches; we explore varied distributions in \S\ref{subsec:eval:compiler}.
Figure~\ref{fig:e2e} presents the average core usage for each application. 
For baselines that fail to meet the target even with maximum resources, we report their packet drop rate in place of a core count.
As shown, \app consistently using the fewest cores.
To understand the sources of this improvement, we now take a deeper look at two representative case studies.

\vspace{-0.5\baselineskip}
\subsubsection{Case Study: NATLB}
\label{subsubsec:eval:natlb}

We first detail NATLB's performance, which is dominated by contention on its shared load-tracking state. 
To isolate this effect, we use the default placement for the \textit{ConnTable} (DRAM on Agilio, eSwitch on BlueField 3) without software caching,
focusing analysis on the frequent, synchronized updates to the \textit{MinLoad} counters for the 256 server groups. 
To avoid cross-device locking when the SmartNIC is overloaded, excess traffic is redirected to the host/Arm cores at the granularity of these server groups.

\para{DPDK} 
baseline uses RSS for incoming traffic, causing poor scaling due to lock contention from scattered updates across host cores to the shared \textit{MinLoad} counters. 
Throughput is limited (BlueField 3: <11 Mpps, Figure~\ref{fig:natlb:core_thput:bf3}; 
Agilio: 12 Mpps, Figure~\ref{fig:natlb:core_thput:agilio}) with high core usage (BlueField 3: 7 Arm cores; Agilio: 8 host cores) 
and significant packet drops (BlueField 3: 57\%/64\%; Agilio: 18\%, Figure~\ref{fig:e2e}).

\para{Static Offload} 
uses hardware steering where possible to direct packets to FPC/DPA threads for counter updates. 
The resulting massive thread contention on popular \textit{MinLoad} group locks severely limits throughput (BlueField 3: <10 Mpps, Figure~\ref{fig:natlb:core_thput:bf3}; 
Agilio: 7 Mpps, Figure~\ref{fig:natlb:core_thput:agilio}), leading to high packet drop rates as the NIC is easily overwhelmed.

\para{Hybrid Offload} 
combines an initial NIC offload with redirecting entire \textit{MinLoad} groups to the host/Arm upon saturation.
This improves locality over DPDK, increasing throughput (BlueField 3: 25 Mpps, Figure~\ref{fig:natlb:core_thput:bf3}; 
Agilio: 13 Mpps, Figure~\ref{fig:natlb:core_thput:agilio}), but still requires substantial host/Arm resources (3 cores on both platforms) 
and drops packets under high load (BlueField 3: 17\%; Agilio: 14\%, Figure~\ref{fig:e2e}).
\revtag{R14}\revised{Whole-group redirection avoids cross-device sharing, but host-side RX steering spreads each group's flows across host/Arm cores contending on the same \textit{MinLoad} lock.
The throughput ceiling causes drops under excess load despite spare cores.}

\para{\app}
\revised{uses model-guided thread allocation and packet steering to manage lock contention on the NIC.}
On BlueField 3, it partitions DPA threads into optimally sized groups (predicted 17, empirical optimum 16, Figure~\ref{fig:natlb:thread_thput:bf3}) using RX steering, 
drastically reducing lock contention and enabling 20 Mpps DPA throughput (Figure~\ref{fig:natlb:thread_table:bf3}). 
Combined with Arm, it reaches 33 Mpps (Figure~\ref{fig:natlb:core_thput:bf3}). 
On Agilio, which lacks fine-grained hardware steering, \app's compiler generates a software dispatcher that steers packets to a dedicated subset of FPC threads, 
achieving a similar contention management effect and improving throughput to 9.5 Mpps (Figure~\ref{fig:natlb:core_thput:agilio}), over Static Offload's 7 Mpps. 
On both platforms, \app uses fewer host/Arm cores than Hybrid Offload (Figure~\ref{fig:e2e}).
Details of the steering implementations are in Appendix~\ref{subsec:backend:steer}.

\para{Latency and Thread Tuning.} 
Reduced contention generally yields lower latency for \app vs baselines (Figure~\ref{fig:natlb:lat_thput:bf3}, \ref{fig:natlb:lat_thput:agilio}). 
Figure~\ref{fig:natlb:thread_table:bf3} shows \app's efficient DPA thread usage by mapping workload carefully based on $N_{critical}$ compared to naive round-robin assignment ("Vanilla") on BlueField 3.
A similar effect is observed on Agilio with the software dispatcher.

\subsubsection{Case Study: McRouter}
\label{subsubsec:eval:mcrouter}

We next evaluate McRouter, focusing on the performance impact of state placement as the hash table size varies 
(up to 1GB Agilio, 16GB BlueField 3 due to DRAM capability limits).
To isolate the impact of state placement, 
we use a key hash embedded in the source UDP port~\cite{lim2014mica} to eliminate lock contention via hardware steering (on host/BlueField 3) 
or mitigate it with hardware atomic engines (on Agilio).

\para{DPDK} baseline runs on host/Arm cores using RSS to avoid atomics. 
However, performance still degrades as the hash table size increases. 
On BlueField 3, it drops 54\% of packets at a 16GB table size vs. the 30 Mpps target (Figure~\ref{fig:e2e} right).

\para{Static Offload} offloads the entire hash table to the NIC. 
Performance drops noticeably on both platforms when the table exceeds the on-chip SRAM capacity 
(512MB on BlueField 3, Figure~\ref{fig:mcrouter_thput:bf3}; 4MB on Agilio, Figure~\ref{fig:mcrouter_thput:agilio}).

\para{Hybrid Offload} yields mixed results. 
On BlueField 3, its throughput approaches the sum of the DPA and Arm capacities, but at the cost of more Arm resources (7 cores, Figure~\ref{fig:e2e} right).
On Agilio, it underperforms DPDK for large tables; the mandatory FPC re-processing of host-handled packets creates a computational bottleneck that outweighs the benefit of offloading lookups to off-chip DRAM (Figure~\ref{fig:mcrouter_thput:agilio}).

\para{\app} optimizes McRouter per platform based on its performance model. 
On BlueField 3, predicting the DPA window overhead (red line, Figure~\ref{fig:mcrouter_thput:bf3}), 
it generates a software cache to keep hot hash table entries in fast on-chip SRAM, steering overflow requests to Arm-shared DRAM. 
This achieves the highest throughput (37 Mpps at 16GB) with minimal Arm cores (Figure~\ref{fig:e2e} right). 
On Agilio, acknowledging the SRAM/DRAM performance gap, \app prioritizes placing hot entries in scarce SRAM. 
While its throughput does not beat DPDK due to the FPC re-processing bottleneck on host interaction (Figure~\ref{fig:mcrouter_thput:agilio}), 
\app's SRAM-aware placement drastically reduces host CPU usage vs. Hybrid Offload (5 fewer cores, Figure~\ref{fig:e2e} left), 
showing efficient on-chip resource use.

{
\begin{figure}[!tp]
\setlength{\abovecaptionskip}{8pt}
\setlength{\belowcaptionskip}{-8pt}
\centering
\minipage{0.44\textwidth}
    \begin{subfigure}{0.49\textwidth}
        \includegraphics[width=\linewidth]{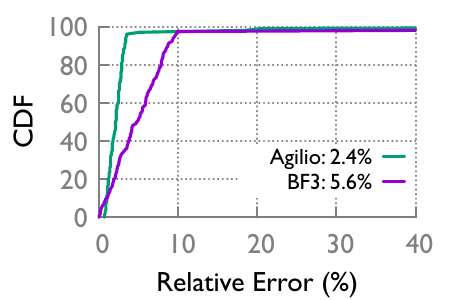}
        \vspace{-18pt}
        \caption{Network functions.}\label{fig:predict-nf}
    \end{subfigure}
    \begin{subfigure}{0.49\textwidth}
        \includegraphics[width=\linewidth]{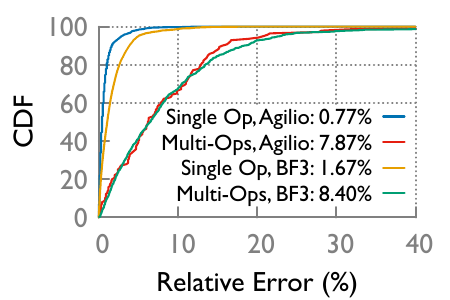}
        \vspace{-18pt}
        \caption{Memory operations.}\label{fig:predict-mem}
    \end{subfigure}
    \hspace*{\fill}
    \caption{CDF of performance model's accuracy.}
    \label{fig:predict-perf}
\endminipage
\vspace{-4pt}
\end{figure}
}

\vspace{-0.5\baselineskip}
\subsection{Compile-Time Adaptability}
\label{subsec:eval:compiler}
We evaluate \app's compile-time adaptability on Agilio with varying constraints and traffic patterns. 
To quantify plan effectiveness, we use the offloading ratio: 
the percentage of the total computational workload (defined in \S\ref{subsec:synthesizer}) that is successfully handled on the SmartNIC.
During allocation, \app seeks to maximize this ratio by adjusting its static memory placement for NATLB and Service Chain workloads.

\para{Impact of fast memory size.}
Simulating NATLB processing 100K flows with varied SRAM (Figure~\ref{fig:sensitivity-agilio:mem_size}). 
With 0 SRAM, 70\% of the computation is offloaded to DRAM. 
As SRAM increases, \app prioritizes the frequently-accessed \textit{MinLoad} counters. 
Subsequently, as more SRAM becomes available, Vela allocates it to the \textit{ConnTable}'s software cache, enabling more hot NAT entries to be promoted at runtime
and achieving near 100\% offload when SRAM suffices for all states.

\para{Impact of traffic load.}
\app adapts allocation to input load changes (varied via 64B packet ratio, higher ratio implies higher Mpps). 
Using its performance model to predict NIC capacity, \app reduces the computation offloaded if the estimated input load exceeds this limit, preventing overload. 
Figure~\ref{fig:sensitivity-agilio:traffic} confirms this: the offloading ratio stays high until the input load surpasses the modeled NIC processing ceiling, 
after which it decreases as computation shifts to the host.

\para{Impact of intra-NF traffic distribution.}
Simulating varied NATLB execution paths (NAT-only vs. NAT+LB traffic), \app adapts SRAM allocation based on the dominant path. 
As shown in Figure~\ref{fig:sensitivity-agilio:state}, when fewer packets require the LB step, 
\app assigns less SRAM to the \textit{MinLoad} counters and more to \textit{ConnTable}'s software cache.
This optimizes SRAM usage for the prevalent workload and maintaining high offloading.

\para{Impact of inter-NF traffic distribution.}
For the concurrent Service Chain, 
\app jointly optimizes allocation considering varying traffic splits between its branches (Figure~\ref{fig:sensitivity-agilio:ratio}). 
\app plans optimized for narrow profiles (A: 10\% NATLB, B: 90\% NATLB) degrade significantly when splits change. 
However, using a broader profile yields \app's robust "Vela Optimal" plan, which balances SRAM allocation between branches, 
sustains high offloading across splits, and outperforms narrowly optimized plans and Hybrid Offload.

\para{Accuracy of performance model.}
\label{subsubsec:eval:performance}
We evaluated \app's performance model by comparing predicted versus measured throughput on Agilio and BlueField-3. 
Tests included the applications from Figure~\ref{fig:e2e} under randomized traffic (flow counts, sizes) 
and synthesized microbenchmarks with diverse memory operations (atomic/bulk, read/write, varying intensity and memory sizes). 
Figure~\ref{fig:predict-perf} presents the CDF of the relative prediction error. 
The model achieves good accuracy, with average relative errors generally below 10\% across NFs and memory operations on both platforms. 
Accuracy is slightly lower on Agilio possibly due to its more explicit memory interface, 
while BlueField 3 shows slightly higher error, 
particularly for operations involving external memory where caching effects can introduce more variability. 

\vspace{-0.5\baselineskip}
\subsection{Runtime Adaptability}
\vspace{-0.2\baselineskip}
\label{subsec:eval:runtime}

{
\begin{figure}[!tp]
\setlength{\abovecaptionskip}{8pt}
\setlength{\belowcaptionskip}{-8pt}
\centering
\minipage{0.5\textwidth}
\centering
  \begin{subfigure}{0.49\textwidth}
    \includegraphics[width=\linewidth]{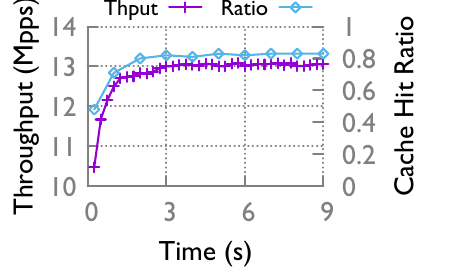}
    \vspace{-18pt}
    \caption{Agilio.}
    \label{fig:agilio_cache}
  \end{subfigure}%
  \begin{subfigure}{0.49\textwidth}
    \includegraphics[width=\linewidth]{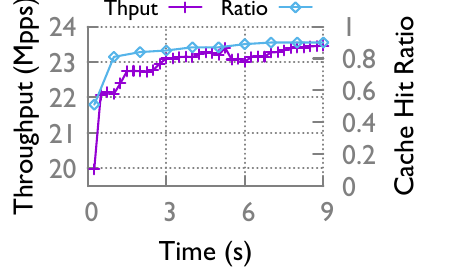}
    \vspace{-18pt}
    \caption{BlueField 3.}
    \label{fig:bf3_cache}
  \end{subfigure}
\endminipage
\caption{Runtime state caching on Agilio and BlueField 3.}
\label{fig:state_cache}
\vspace{-8pt}
\end{figure}
}

We evaluate \app's runtime state caching using McRouter with our skewed Zipf (0.99) request workload. 
We allocate all available on-chip SRAM for the software cache ($\sim$4MB on Agilio, $\sim$512MB on BlueField 3) against a much larger total hash table residing in DRAM (64MB and 2048MB, respectively).
We pre-populate all keys and begin the experiment with a cold cache.
Figure~\ref{fig:state_cache} shows that the runtime quickly populates the cache with hot entries, reaching a stable, warm state in approximately 3 seconds. 
The final cache hit ratio reaches 0.83 on Agilio and 0.90 on BlueField 3. 
\revtag{16}\revised{With this fixed allocation, runtime caching improves throughput over DRAM-only placement} (i.e., the Static Offload baseline)
by 20.0\% on Agilio (from 10.9 Mpps to 13.1 Mpps) and 8.1\% on BlueField 3 (from 21.7 Mpps to 23.5 Mpps).

%% file: 8-discussion.tex
\vspace{-0.5\baselineskip}
\section{Discussion}
\label{sec:discussion}

\para{Generalizability of the performance model.} 
\revtag{R5}\revised{\app's performance model separates memory engines from the memory resources they access (Figure~\ref{fig:memArch}).
A new memory type, such as High Bandwidth Memory (HBM), is incorporated by profiling its supported engine-resource access paths through one-time offline microbenchmarks (\S\ref{subsec:perform:roofline}).
The resulting \linebreak curves capture its latency and throughput characteristics using the same parameterized model (Equation~\ref{eq:mem_roofline}).
Replacing DRAM with HBM therefore changes the hardware-specific profiles and capacity constraints, but not the Predictor's composition method or the Synthesizer's allocation logic.}
We argue that this one-time, upfront modeling effort is a worthwhile investment.
At the scale of modern data centers, where a single SmartNIC model may be deployed by the millions, 
the benefits of predictable performance and optimized resource allocation justify platform-specific modeling.

\para{Comparison with the state-of-the-art.}
Existing SmartNIC optimization tools are largely complementary to \app, as they focus on different architectural components or optimization layers. 
Frameworks like Pipeleon~\cite{xing2023pipeleon} and Yala~\cite{wu2025yala}, for instance, model P4 pipelines and hardware accelerators. 
\app's detailed analysis of the general-purpose memory subsystem and software lock contention can be integrated to provide a more holistic performance view. 
Similarly, \app's compile-time analysis can generate an optimal static plan for runtime schedulers like iPipe~\cite{ipipe} to manage.

\app also complements tools with different optimization targets. 
Clara~\cite{DBLP:conf/sosp/QiuXHKLNC21}, which models latency, could be used alongside \app in a co-optimizer to meet both throughput and latency goals. 
\revtag{R15}\revised{Alkali~\cite{lin2025alkali} optimizes pipeline partitioning and state placement across heterogeneous SmartNICs.
It partitions mutable table state by key for lock-free parallel execution across cores, leaving memory contention and locking costs outside its performance model.
\app instead models memory-tier contention and software-lock throughput ceilings to jointly determine state placement and useful thread parallelism for shared-state processing.
Alkali's high-level abstractions could serve as inputs to \app, whose contention-aware performance estimates could in turn guide Alkali's parallelization decisions.}

\para{\revised{Alternative Synchronization and Execution Models.}}
\revised{\app\revtag{R8} models hardware atomics and software spinlocks in a run-to-completion (RTC) datapath.
Other synchronization mechanisms offer complementary benefits.
Read-copy-update (RCU) permits readers to access published versions without reader--writer locking~\cite{linux_rcu}, while seqlocks validate reads and retry on overlapping updates~\cite{linux_seqlock}.
These mechanisms reduce reader synchronization costs; however, concurrent multi-instruction, conditional updates across variables, such as NATLB's \{\textit{MinLoad}, \textit{MinDip}\} block (Figure~\ref{fig:dep_graph}), still require writer coordination.}

\revised{For updates, queue-based locks and synchronous delegation can reduce contention overhead~\cite{mcs1991,qd2014} while retaining RTC processing.
The worker still waits for lock acquisition or update completion, with potential savings determined by queue placement, polling, notification, and handoff costs.
Realizing these benefits requires adapting to the target: Agilio FPCs lack private data caches~\cite{krude2021throughput}, so CPU-style cache-local polling does not transfer directly, and queue management must share limited code-store and local-memory capacity with NF logic~\cite{deepmatch}.
BlueField DPA offers caches and event-driven execution, with explicit memory ordering needed for queue publication and completion notification~\cite{flexio}.}

\revised{Asynchronous delegation can further overlap updates with subsequent work~\cite{qd2014}.
An RTC worker could finish its packet and accept the next before the update completes, provided this preserves the NF's required update ordering and visibility.
Supporting this in \app would require establishing when such deferral is valid, alongside resource management for request buffering, completion tracking, and backpressure.}

\revised{Coroutines, task-based execution, and continuation passing offer another way to use waiting time by suspending unfinished packets to process others.
This extends the execution model beyond RTC and requires packet-level scheduling and context management.
On Agilio, hardware context switching operates on a fixed set of FPC threads~\cite{deepmatch}; multiplexing unfinished packets within a thread would add software scheduling and context storage.
DPA's cooperative rescheduling provides a useful building block, though packet contexts must be preserved explicitly because handler stacks may not persist across invocations~\cite{flexio}.
Extending \app to these models would require accounting for their scheduling, context-storage, and synchronization costs.}

\para{Handling Workload Pattern Shifts.}
\revtag{R11}
\revised{
A significant and persistent workload shift may call for a substantially different memory layout, requiring the NF to be recompiled and its firmware reloaded.
Because SmartNIC memory layouts are fixed at compile time, the old and new plans cannot coexist on the same device; installing new firmware inherently interrupts the running NIC.
Traffic must therefore be temporarily redirected to host CPUs or a standby NIC during the update, requiring additional processing capacity.
Compounding this disruption, closed-source vendor compilation and firmware loading are minute-scale operations, making end-to-end reconfiguration both slow and resource-intensive.

\app therefore seeks to reduce how often full reconfiguration is needed instead of relying on frequent firmware updates to track workload fluctuations.
\app's analysis and synthesis can generate a new plan within seconds.
To reduce the frequency of minute-scale redeployments, the Synthesizer can optimize over a longer-horizon trace or multiple anticipated workload profiles.
The resulting plan trades slight per-profile optimality for robustness across the expected range.
As shown in our Service Chain evaluation (Figure~\ref{fig:sensitivity-agilio:ratio}), a plan generated from an aggregated traffic profile maintains high efficiency across widely varying traffic splits.
Firmware redeployment is thus reserved for sustained shifts outside the range covered by the robust plan.
}

%% file: 10-conclusion.tex
\vspace{-0.5\baselineskip}
\section{Conclusion}

This paper presented \app, a predictive compiler framework that addresses the challenge of 
efficiently offloading stateful NFs to multi-threaded SoC SmartNICs.
At its core is a novel, state-centric analytical model that composes the performance of primitive operations 
to estimate the throughput ceiling of a given resource allocation plan,
accounting for both memory system bottlenecks and software lock contention. 
Evaluation on two SmartNICs shows that \app generates efficient, 
plans that significantly reduce host core consumption compared to traditional approaches.

%% file: appendix.tex
\appendix

\smallskip
\smallskip

\begin{center}
{\bf \large APPENDIX}
\end{center}

\section{Netronome Agilio SmartNIC's FPC Processing Stages and Profiling}
\label{sec:appendix:profile}

In Agilio SmartNICs, packets are processed at the Flow Processing Cores (FPCs).
In this section, we first introduce how FPCs process packets
(\S\ref{subsec:perform:arch}).  Next, we run a series of profilings to
demonstrate the performance characteristics of FPC-local operations
(\S\ref{subsec:perform:microbench}).

\subsection{FPCs' Architecture \& Processing Flow}
\label{subsec:perform:arch}

{
\centering
\setlength{\abovecaptionskip}{-.1pt}
\begin{figure}[t] \centering
\includegraphics[width=0.9\columnwidth]{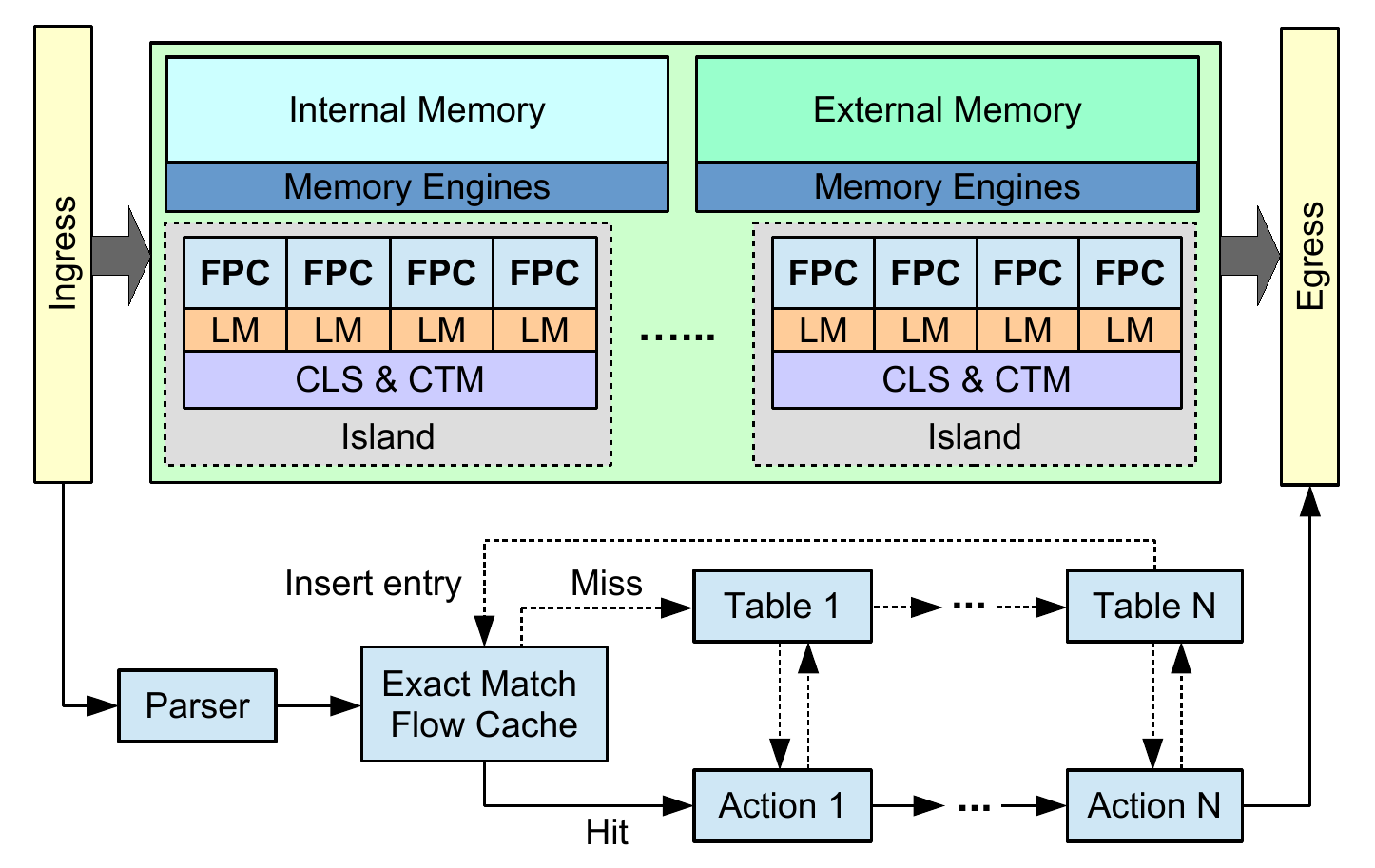}
\caption{Agilio Architecture and FPC processing workflow.}
\label{fig:nfp_arch}
\end{figure}
}

{
\begin{table}[tp!]
{
    \centering \scriptsize \renewcommand{\arraystretch}{1.2}
    \begin{tabular}{|c|c|c|c|}
    \hline
    {\bf Memory} & {\bf Scope} & {\bf Size} & {\bf Latency (cycles)} \\ \hline\hline
    Instruction Memory (IM) & Core & 8K & N/A \\ \hline
    Local Memory (LM) & Core & 4KB & 1-3 \\ \hline
    Cluster Local Scratch (CLS) & Island\tnote{1} & 64KB & 20-50 \\ \hline
    Cluster Target Memory (CTM) & Island & 256KB & 50-100 \\ \hline
    Internal Memory (IMEM/SRAM) & Global & 4MB & 150-250 \\ \hline
    External Memory (EMEM/DRAM) & Global & 2GB & 150-500 \\ \hline
    \end{tabular}
   \vspace{2\baselineskip}
    \caption{FPC memory hierarchy.}
    \label{tab:agilio_memory}
    }
\end{table}
}

{
\begin{figure*}[!t]
\setlength{\abovecaptionskip}{2pt}
\minipage{\textwidth}
\centering
\begin{subfigure}{0.33\textwidth}
  \includegraphics[width=\linewidth]{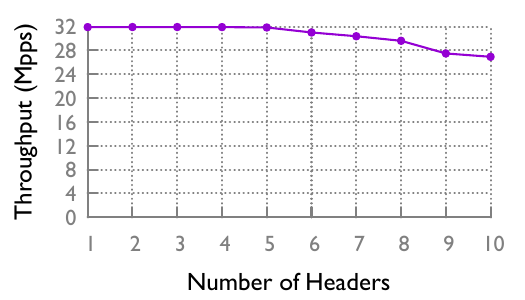}
  \caption{Impact of number of headers.}
  \label{fig:hdr_num}
\end{subfigure}
\begin{subfigure}{0.33\textwidth}
  \includegraphics[width=\linewidth]{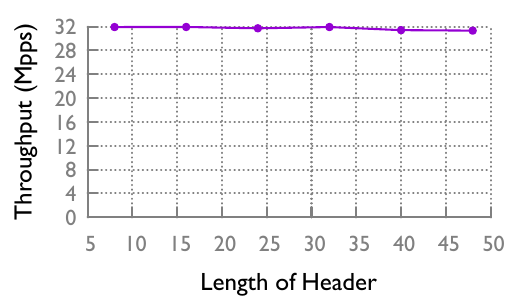}
  \caption{Impact of header field length.}
  \label{fig:hdr_len}
\end{subfigure}
\begin{subfigure}{0.33\textwidth}
  \includegraphics[width=\linewidth]{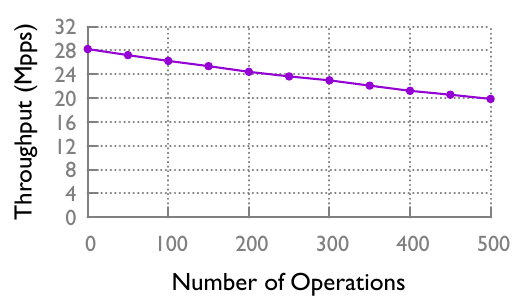}
  \caption{Impact of number of arithmetic operations.}
  \label{fig:alu}
\end{subfigure}
\endminipage
\caption{Throughput profiling of P4 parser and arithmetic operations on Netronome SmartNIC.}
\end{figure*}
}

An FPC is a 32-bit RISC core with up to 8 thread contexts, with at most one thread
executing at a time.  FPCs use the run-to-completion architecture, rather than the
pipeline architecture used by programmable switching ASICs.  Each FPC thread
executes the compiled instructions sequentially and accepts the next packet
after processing the current one.
Figure~\ref{fig:nfp_arch} shows the Agilio architecture and how FPCs process packets in Agilio SmartNICs.

\para{FPC memory hierarchy.}
Shown in Table~\ref{tab:agilio_memory}, FPC interacts with instruction memory and five tiers of data memory, each with
its own scope, size, and access latency. 

The two globally-scoped memory tiers available to all FPCs are Internal Memory (IMEM) and External Memory (EMEM), 
which correspond to the on-chip SRAM and off-chip DRAM, respectively. 
We use these global memories to store variables shared by all FPCs, such as locks, counters, and global states. 
FPCs rely on Memory Engines (MEs) to perform read and write operations. 
There are two types of MEs, bulk and atomic. 
Crucially, IMEM and EMEM are serviced by their own dedicated MEs and do not overlap.

\para{FPC programming language.}
FPCs support both P4 and Micro-C languages.
An FPC program is similar to a typical P4 program.
Programmers can use the P4 language to define header, parser, and match fields of the tables.
As for  actions, programmers usually use the Micro-C language for better flexibility.

\para{FPC processing flow.}
Shown in Figure~\ref{fig:nfp_arch}, when a packet arrives at the ingress,
an FPC thread executes the following steps:

\squishlist
\item Parse packet headers stored in CTM, and extract header fields.
\item Look up Exact Match Flow Cache (EMFC), which sits in EMEM and stores all visited flows.
\item If the packet belongs to a new flow, the FPC executes the entire program, \ie, 
looks up the tables and executes the matched actions.
At the same time, the FPC records the matched entry in each table (value of matched fields,
action ID and corresponding parameters) and pushes them into the EMFC.
\item If the packet header fields are found in the EMFC, 
the FPC skips all table lookups and executes only the recorded actions.
\item Push the packet into the egress and process the next packet.
\squishend

\subsection{Profiling Results and Analysis}
\label{subsec:perform:microbench}

We ran a series of profilings to quantify different factors' impact on packet processing throughput, as summarized in Table~\ref{tab:fpc_factors}. Across all
profilings, we injected the smallest-sized packets (no smaller than 64 bytes)
and recorded the maximum packet processing throughput that Agilio can process
with no packet drop. 

{
\begin{table}[tp!]
{
\makebox[\columnwidth][c] {
\begin{threeparttable}
    \centering \scriptsize \renewcommand{\arraystretch}{1.2}
    \begin{tabular}{|c|c|c|}
    \hline
    \textbf{Processing Step / Operation} & \textbf{Impact Factors}   & \textbf{Profile Result} \\ \hline\hline
    \multirow{3}{*}{Header Parsing} 
    & Number of headers & Figure~\ref{fig:hdr_num} \\ \cline{2-3} 
    & Header length & Figure~\ref{fig:hdr_len} \\ \hline
    \multirow{4}{*}{Table Lookup\tnote{1}} 
    & Number of matched fields & \multirow{4}{*}{Figure~\ref{tab:emfc}} \\ \cline{2-2} 
    & Width of matched fields  & \\ \cline{2-2} 
    & Number of tables & \\ \cline{1-2} 
    Exact Match Flow Cache & Width of EMFC key & \\ \hline
    Arithmetic Operation   & Number of operations & Figure~\ref{fig:alu} \\ \hline
    \end{tabular}
    \begin{tablenotes}
        \item[1] We disabled EMFC in this microbenchmark.
    \end{tablenotes}
    \vspace{2\baselineskip}
    \caption{Performance profiling of FPC operations.}
    \label{tab:fpc_factors}
\end{threeparttable}}}
\end{table}
}

\para{Profiling baseline.}
In the baseline, FPCs only forward packets from ingress to egress. 
Here, we measure a total throughput of \textbf{31.91} Mpps.
This shows the best-case throughput, which 
is bottlenecked by other components in Agilio.

\para{Header parsing.} 
Packet headers are stored in FPC's CTM.  Header parsing is the first step in
packet processing.  FPCs receive the pointer to the packet header, executes the
parsing state machine, and records each header field's offset in the LM.  The
parsing procedure is performed locally in each FPC. The results are shown in
Figure~\ref{fig:hdr_num} and Figure~\ref{fig:hdr_len}. We can see that the
forwarding performance is unaffected until the FPC parses more than 8 headers.

\begin{figure*}[tp]
    \centering
    \begin{minipage}{0.33\textwidth}
        \centering
        \scriptsize
        \renewcommand{\arraystretch}{1.2}
        \begin{threeparttable}
            \centering \scriptsize \renewcommand{\arraystretch}{1.2}
            \begin{tabular}{|c|c|c|c|}
              \hline
              \multirow{2}{*}{Table \#} & \multirow{2}{*}{Field \#} & \multicolumn{2}{c|}{Throughput (Mpps)} \\ 
              \cline{3-4}
              & & Disable EMFC & Enable EMFC \\ 
              \hline\hline
              1 & 1 & 25.43 & 31.43 \\ \hline
              1 & 2 & 15.54 & 31.37 \\ \hline
              2 & 1 & 16.64 & 29.91 \\ \hline
              1 & 4 & 5.62  & 31.30 \\ \hline
              4 & 1 & 9.48  & 29.20 \\ \hline
              2 & 2 & 10.89 & 29.95 \\ \hline
            \end{tabular}
            \caption{Throughput under table lookup.}
            \label{tab:emfc}
        \end{threeparttable}
    \end{minipage}
    \hfill
    \begin{minipage}{0.65\textwidth}
        \centering
        \begin{subfigure}[b]{0.48\linewidth}
            \includegraphics[width=\linewidth]{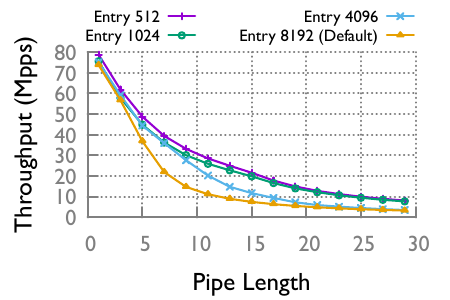}
            \vspace{-16pt}
            \caption{Impact of number of entries.}
            \label{fig:pipe_len}
        \end{subfigure}
        \hfill
        \begin{subfigure}[b]{0.48\linewidth}
            \includegraphics[width=\linewidth]{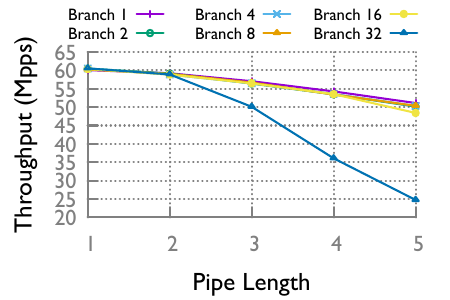}
            \vspace{-16pt}
            \caption{Impact of number of branches.}
            \label{fig:pipe_width}
        \end{subfigure}
        \vspace{-8pt}
        \caption{Throughput under varying pipe length and branches.}
        \label{fig:bf3_eswitch}
    \end{minipage}
\end{figure*}

\para{Table lookup \& action.}
The table entries are stored in the EMEM.  FPCs apply
the DCFL~\cite{taylor2005scalable} algorithm to look up tables.  DCFL is a decomposed
searching algorithm that searches each field of a table separately and
aggregates search results to return the final match result.  In FPCs,
the per-field searching is accelerated by a lookup engine.  However, because of the
high EMEM access latency, the DCFL algorithm still performs poorly.  FPCs spend most
of the processing cycles in table actions.  In the actions, FPCs perform
diverse operations, \eg, arithmetic/bit operations, read/write different
memories, acquire locks, \etc

Shown in Table~\ref{tab:emfc}, we can see that the performance drops quickly as
the number of tables increases from 1 to 4. When the number of tables is fixed to 1,
the number of matched fields also impacts the performance.
This is caused by the DCFL algorithm adopted by Agilio.

\para{Exact match flow cache.}
To accelerate the table lookup process, the FPC employs a special data structure
called Exact Match Flow Cache (EMFC)~\cite{agilio_ovs_software}.  EMFC is an
exact match hash table stored in the EMEM.  Its match field is the concatenation
of all tables in the program.  EMFC's key is composed during packet header
parsing. When a packet from a new flow arrives, EMFC caches the matched
entries of all tables and corresponding actions along with the parameters. 
Later, when subsequent packets arrive, the FPC can skip all table lookups and only
execute the actions recorded in EMFC.
EMFC has a size limit of 2 million.

We compare the packet processing throughput before and after EMFC is enabled, as shown in
Table~\ref{tab:emfc}. We can see that the EMFC can greatly accelerate the packet
processing. And the performance is not affected by the number of tables and the
number of fields.

\para{Arithemetic Operation.}
Shown in Figure~\ref{fig:alu}, arithmetic operations can be performed efficiently. We execute operations of addition, subtraction, and bitwise XOR, AND, and OR in a recurring sequence. We only see major performance drop after executing 200 individual operations.

\section{BlueField 3 SmartNIC's eSwitch Profiling}
\label{appendix:bf2_profiling}

eSwitch is BlueField 3 SmartNIC's solution to accelerate table lookups. eSwitch allows
the programmers to define a series of lookup tables. Due to hardware limitation, eSwitch
only supports pre-defined fields and actions. eSwitch supports two modes, Control Pipe and Miss Pipe, in flexibly combine tables to form a pipeline. In Control Pipe, programmers can define the priority of the entries in the table and jump to different tables based on the matching result.
Miss Pipe, on the other hand, only looks up the next table when the previous one misses.
We conducted experiments to characterize the impact of table combinations on eSwitch performance.

To assess the impact of pipeline length on throughput, we sequentially chained various numbers of tables using the Miss pipe.
Each pipeline contained a fixed number of flow entries. 
Figure~\ref{fig:pipe_len} shows that accommodating more flow entries in a long matching pipeline leads to significant throughput drops. 
Conversely, packing more entries into a single pipe enhances throughput.

To accelerate table matching with a large number of flow entries (> 1M), we distributed flow entries equally among multiple branches.
A control pipe was used to partition flows into non-overlapping matching spaces, such as matching IP addresses with an N-bit mask. 
Each subspace was handled by a chain of pipes, each containing 8,192 entries. 
Figure~\ref{fig:pipe_width} demonstrates that increasing the number of branches (subspaces) results in higher throughput. 
For example, using 32 branches achieves the same throughput as using 16 branches but accommodates up to twice 
as many flow entries when the pipe length does not exceed two. 
However, as the pipe length increases, higher branch counts lead to more severe throughput degradation. 
Additionally, we examined the effect of various table lookup parameters on forwarding throughput, 
finding that the forwarding throughput remains consistent regardless of the width of matched fields when increasing the total number of matched fields from 1 to 16. 
Given eSwitch's predictability, we rely on throughput results from micro-benchmarking experiments to guide its performance prediction.

{
\begin{figure}[t] \centering
\includegraphics[width=0.47\textwidth]{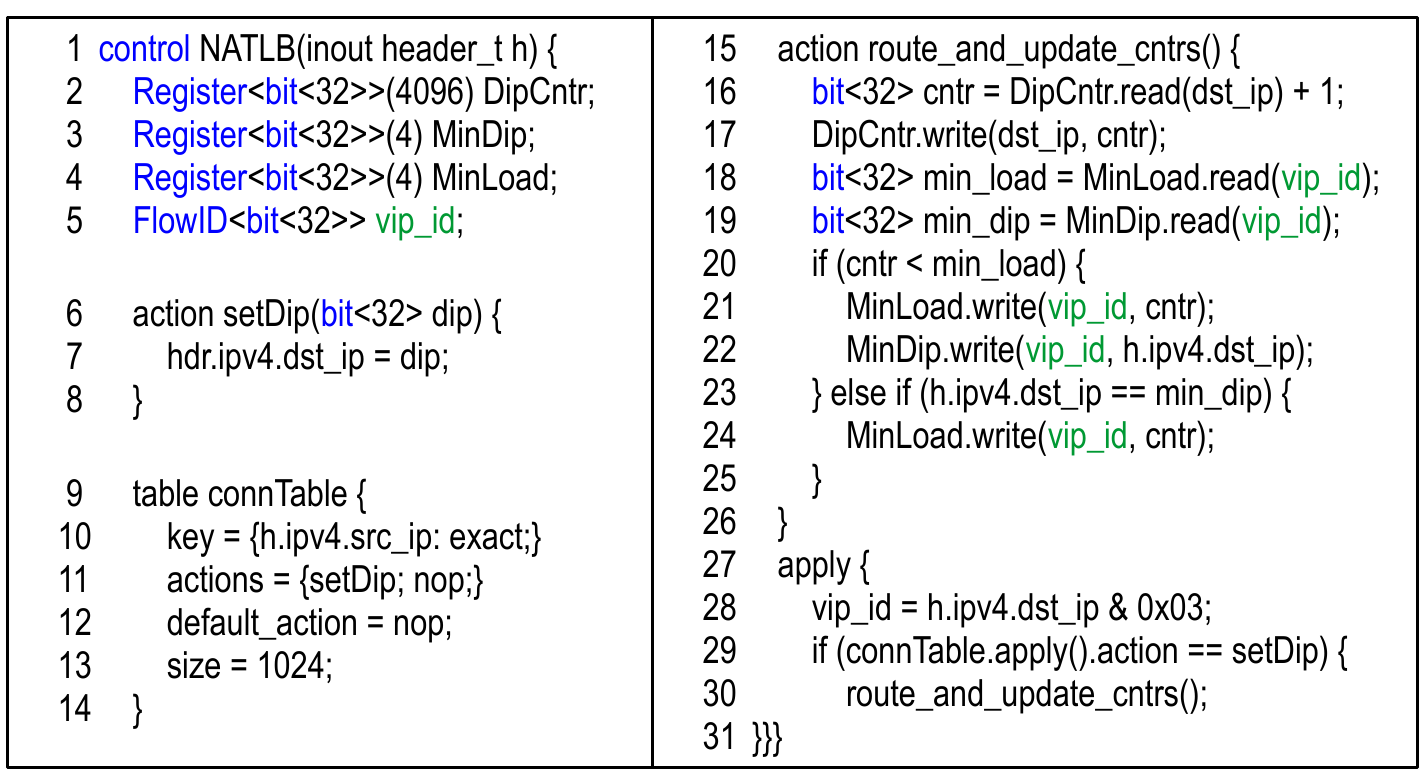}
		\caption{The P4 program of NATLB.}
\label{fig:ir_example}
\end{figure}
}

\section{Code Analysis and Execution Example}
\label{appendix:ir}

The generic analysis process described in \S\ref{subsec:analyser} can be instantiated for specific high-level languages like P4. 
This appendix details how the Analyzer extracts the state-flow representation from the P4 implementation of the NATLB example (Figure~\ref{fig:motiv_example}).

\subsection{State-Flow Representation in P4}
\label{subsec:frontend:representation}

{
\begin{figure}[t] \centering
\includegraphics[width=0.44\textwidth]{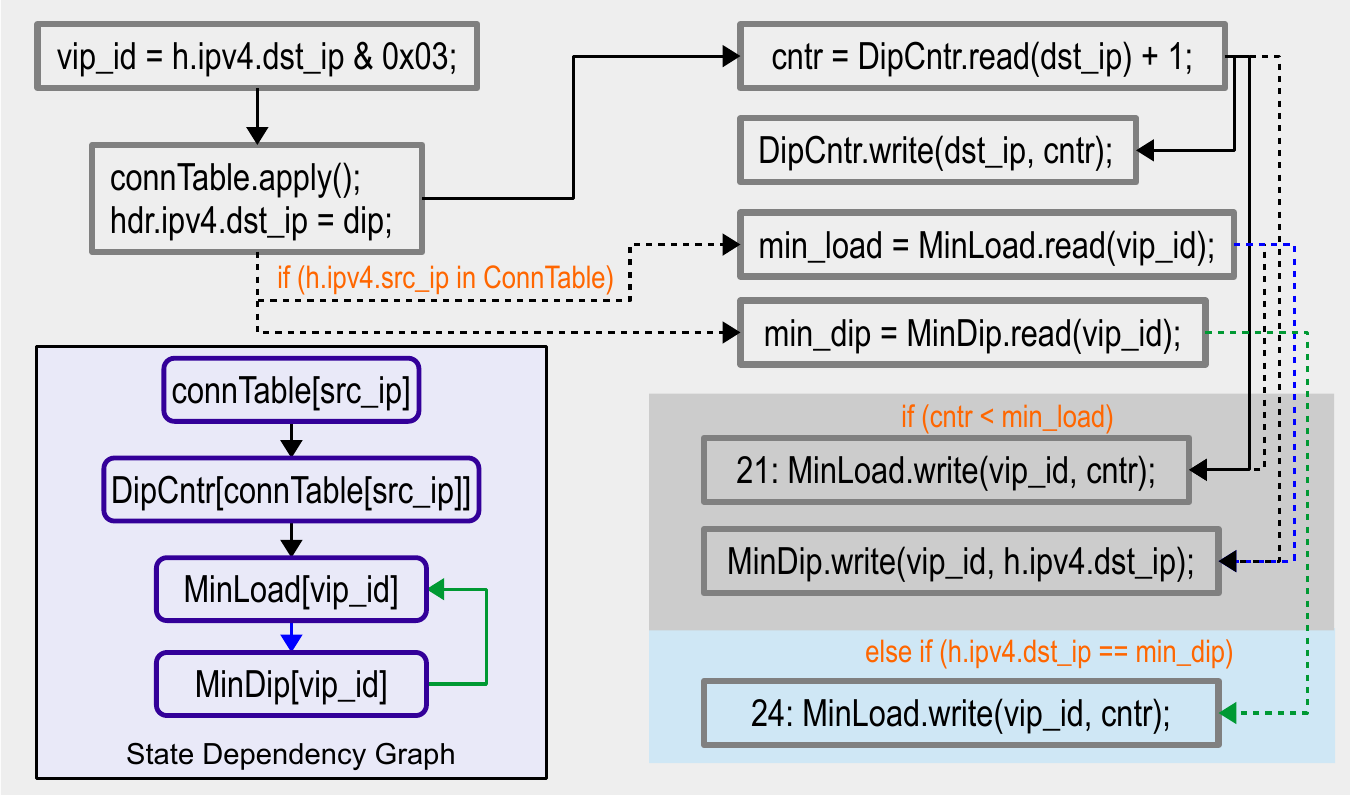}
  \caption{NATLB dependency graph illustrating brach conditions (orange), data (solid edges), control (dashed), and extracted state dependencies (subgraph).}
\label{fig:dep_graph}
\end{figure}
}

The Analyzer leverages P4's structured nature to map its constructs to our abstract IR.
It generates a dependency graph (Figure~\ref{fig:dep_graph}) where nodes are instructions and edges represent data and control flow. 

\noindent\textbf{State Block:}
P4's explicit \textit{table} and \textit{register} declarations map directly to stateful objects. 
The Analyzer parses the dependency graph to identify cyclic read-after-write dependencies. 
As shown, the conditional updates between \textit{MinLoad} and \textit{MinDip} form such a cycle, causing the Analyzer to group them into a single atomic state block. 
\textit{ConnTable} and \textit{DipCntr} form their own distinct blocks.

\noindent\textbf{Flow Group Identifier:}
To make the link between traffic and state explicit, we introduce a custom FlowGroupID annotation type. 
Programmers use this to mark which header fields or metadata serve as the index for a given state. 
In Figure~\ref{fig:ir_example} (Line 5 \& 28), \textit{vip\_id} is declared as a FlowGroupID, 
derived from \textit{h.ipv4.dst\_ip}. All packets that hash to the same \textit{vip\_id} belong to the same flow group.

\noindent\textbf{State-Flow Mapping:}
By parsing the P4 code, the Analyzer determines that the FlowGroupID \textit{vip\_id} is used as the index for accessing the \{\textit{MinLoad, MinDip}\} state block, 
thus establishing the mapping $\mathcal{T}_{\{MinLoad, MinDip\}}$.

\para{State Block Analysis.} 
The Analyzer identifies the cyclic data dependencies between the read of \textit{MinLoad} (Line 18) influencing the write to \textit{MinDip} (Line 22), 
and the read of \textit{MinDip} (Line 19) influencing the write to \textit{MinLoad} (Line 24). 
This grouping of interdependent operations shown in the state dependency subgraph, conceptually similar to Domino's instruction condensation~\cite{sivaraman16packet}, 
forms the basis for creating the \{\textit{MinLoad, MinDip}\} state block, 
which is marked as requiring a software lock for atomicity.

\vspace{-.5\baselineskip}
\subsection{Enforcing Parallelism via Packet Steering}
\label{subsec:backend:steer}

{
\setlength{\abovecaptionskip}{4pt}
\setlength{\belowcaptionskip}{-4pt}
\begin{figure}[tp] \centering
\includegraphics[width=\columnwidth]{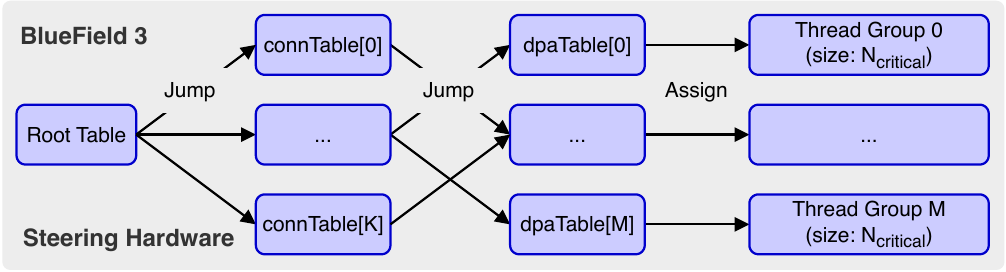}
\includegraphics[width=\columnwidth]{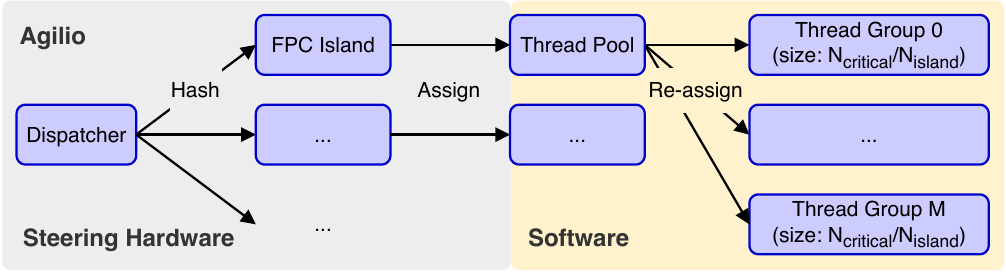}
\caption{Thread group isolation on BlueField 3 and Agilio.}
\label{fig:steer}
\end{figure}
}

The state-flow representation provides the crucial link between application logic and performance, but realizing the model's benefits requires a mechanism to enforce its constraints at runtime. 
Accesses to atomic state blocks like \{\textit{MinLoad, MinDip}\} can cause severe lock contention in a multi-threaded environment.
\app's performance model (\S\ref{subsec:perform:lock}) calculates the optimal number of threads, $N_{critical}$, 
that can service a single lock instance without performance degradation. 
The Synthesizer's task is to generate a configuration that enforces this thread count.

The key mechanism is packet steering: by controlling which threads process packets belonging to a specific flow group ($vip\_id$), 
we can effectively isolate a dedicated set of $N_{critical}$ threads for each contended lock instance. 
Figure~\ref{fig:steer} illustrates how this is achieved on our two distinct hardware platforms.

\para{BlueField 3.}
BlueField 3's flexible RX steering hardware allows for direct, fine-grained control. 
We implement a two-stage steering pipeline. 
An initial \textit{connTable}, matched on packet headers, directs the flow to a secondary \textit{dpaTable}. 
The control plane populates these tables, mapping all flows that share the same \textit{vip\_id} to the same \textit{dpaTable} instance. 
Each \textit{dpaTable} is, in turn, configured to steer its traffic to a dedicated and isolated thread group of size $N_{critical}$. 
This hardware-based approach guarantees that all accesses to a specific \{\textit{MinLoad, MinDip}\} instance are handled exclusively by its assigned thread group, 
precisely enforcing the parallelism constraint predicted by our model.

\para{Agilio.}
Agilio's hardware provides only coarse-grained steering, hashing packets to a large "island" of FPC threads. 
A packet is assigned to any available thread from a shared pool.
To achieve the fine-grained control, \app's compiler generates a software dispatcher that emulates the behavior of BlueField 3's hardware steering.
Inspired by prior work~\cite{shashidhara2022flextoe,lin2025alkali}, we partition the FPC threads within an island into logical groups of size $N_{critical}/N_{island}$, 
where $N_{island}$ is the number of FPC islands. 
When a thread initially receives a packet, it first determines the packet's correct destination thread group based on its \textit{vip\_id}. 
If the current thread already belongs to the target group, it processes the packet directly. 
Otherwise, it re-enqueues the packet's descriptor to a lightweight work queue associated with the correct destination group. 
Because packet headers reside in memory shared across the island, a thread in the destination group can then dequeue the descriptor and process the packet efficiently. 
This software-based steering ensures that accesses to any single locked state are still isolated to a specific thread group, 
effectively enforcing the $N_{critical}$ constraint despite hardware limitations.

\section{Composing Performance Models and\\Prediction Workflow}
\label{subsec:perform:hybrid}

\begin{table}
\centering
\minipage{\linewidth}
\centering \scriptsize \renewcommand{\arraystretch}{1.2}
\begin{tabular}{|c|c|c|}
\hline
\multirow{2}{*}{\textbf{Access Params}} & \multicolumn{2}{c|}{\textbf{Bottleneck}} \\ \cline{2-3}
 & \textbf{Agilio} & \textbf{BlueField 3} \\ \hline
Same $t$, $e$          & Memory engine    & Memory engine \\ \hline
SRAM, different $e$    & The slower op    & \multirow{2}{*}{/} \\ \cline{1-2}
DRAM, different $e$    & Memory bandwidth &  \\ \hline
Different $t$, $e$     & The slower op    & The slower op \\ \hline
\end{tabular}
\caption{Performance bottleneck of hybrid memory access on Agilio and BlueField 3.}
\label{tab:bottleneck}
\endminipage\hfill
\end{table}

\begin{table*}[]
\centering
\scriptsize
\setstretch{1.2}
\begin{tabular}{|c|ccccc|c|c|c|}
    \hline
    \multirow{2}{*}{\textbf{State Block}} & \multicolumn{5}{c|}{\textbf{Access Parameters}} & \multirow{2}{*}{\textbf{Relative Rate}} & \multirow{2}{*}{\textbf{Intensity ($i$)}} & \multirow{2}{*}{\textbf{Aotmicity}} \\ \cline{2-6}
    & \textbf{Tier ($t$)} & \textbf{Op ($o$)} & \textbf{Engine ($e$)} & \textbf{Data Size ($s$)} & \textbf{Array Size ($n$)} & & & \\ \hline\hline
    \multirow{3}{*}{\textit{ConnTable}} & IMEM & Read & Atomic & 4B & 131072 & 1.0 & 1 & \multirow{3}{*}{Mem Engine} \\ \cline{2-8}
     & IMEM & Read & Atomic & 32B & 32768 & 0.8 & 1 & \\ \cline{2-8}
     & EMEM & Read & Atomic & 32B & 98304 & 0.2 & 1 & \\ \hline
    \textit{DipCntr} & IMEM & Inc & Atomic & 4B & 4096 & 1.0 & 1 & Mem Engine \\ \hline
    \multirow{2}{*}{(\textit{MinLoad}, \textit{MinDip})} & IMEM & Read & Bulk & 4B & 256 & 1.0 & 2 & \multirow{2}{*}{SpinLock} \\ \cline{2-8}
     & IMEM & Write & Bulk & 4B & 256 & 1.0 & 2 & \\ \hline
\end{tabular}
\caption{Example state access parameters for throughput prediction.}
\label{tab:state_parameters}
\end{table*}

\begin{figure}[tp]
\centering
\setlength{\belowcaptionskip}{-4pt}
\begin{subfigure}{\linewidth}\centering
  \setlength{\abovecaptionskip}{-0.5pt}
  \includegraphics[width=0.7\linewidth]{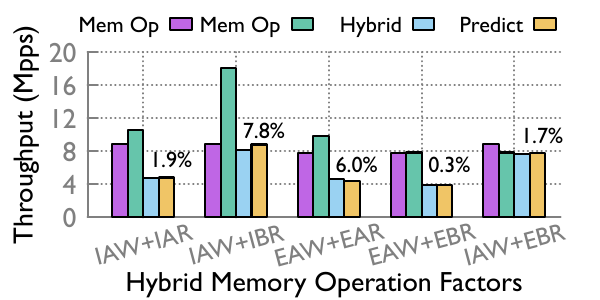}
  \caption{Agilio (Abbreviations: I-IMEM, E-EMEM, A-Atomic, B-Bulk, R-Read, W-Write).}
\end{subfigure}
\begin{subfigure}{\linewidth}\centering
  \setlength{\abovecaptionskip}{-0.5pt}
  \includegraphics[width=0.7\linewidth]{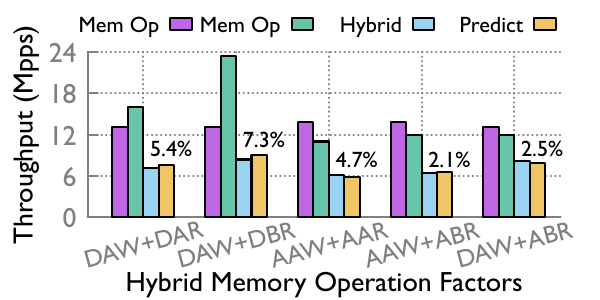}
  \caption{BlueField 3 (Abbreviations: D-SRAM Memory allocated in DPA heap, A-DRAM Memory allocated in ARM huge pages, A-Atomic, B-Bulk, R-Read, W-Write).}
\end{subfigure}
\caption{Performance characteristics of hybrid memory access on Agilio and BlueField 3.}
\label{fig:hybrid}
\end{figure}

This appendix details the methodology for composing primitive performance curves to predict the throughput for a complete, lock-free NF workload, 
and provides a step-by-step example of the full prediction workflow.

\subsection{Modeling Hybrid Lock-Free Accesses}

An NF workload typically consists of a mix of different primitive memory operations. 
These operations may interfere if they share hardware resources. 
Our model composes the individual performance curves of these primitives to derive a single, composite curve for the entire workload, 
from which we calculate the final lock-free throughput ($T_{lockfree}$).

\para{Bottleneck Analysis.} 
The composition rule depends on whether operations share a hardware bottleneck. 
Inspired by Gables~\cite{Gables}, we empirically profiled pairs of operations to identify shared bottlenecks on each platform (summarized in Table~\ref{tab:bottleneck}). 
For example, on Agilio, two different atomic operations on IMEM will share the same memory engine, 
whereas an atomic operation on IMEM and a bulk operation on EMEM will execute on parallel engines.

\para{Deriving a Composite Performance Curve.}
For a set of operations $\{x\}$ that share a memory engine bottleneck, we derive a new, composite performance curve that represents their collective behavior. 
This is a two-step "synthesize-and-fit" process:

\noindent\textbf{Step 1: Synthesize Data Points:} 
We first generate a new set of empirical data points $(i, M_{e}(i))$ for the mixed workload. 
The intuition is to calculate the weighted harmonic mean of the individual throughputs, which correctly models systems where tasks share a common service resource. 
We use the following function to calculate the memory operation throughput $M_e$ at any given test intensity $i$:
\begin{equation}
\footnotesize
    \label{eq:mem_compose}
    M_{e}(i) = \frac{I_{total}}{\sum_{x} \frac{i_x \cdot R_x}{M_x(i)}}.
\end{equation}
Here, $I_{total} = \sum_x (i_x \cdot R_x)$ is the constant total intensity of the actual NF workload, 
$i_x$ is the intensity of primitive operation $x$, $R_x$ is its relative execution rate, and $M_x(i)$ is the throughput of primitive $x$ from its profiled curve at test intensity $i$. 
We evaluate this function for a range of $i$ values (e.g., 1, 2, ..., N) to generate a new set of data points.

\noindent\textbf{Step 2: Fit a New Analytical Curve:} 
We then fit these synthesized data points to our analytical model (Equation~\ref{eq:mem_roofline}) to obtain a new set of composite parameters ($a_{comp}$, $b_{comp}$, $M_{\{roof,comp\}}$). 
This step is crucial because the fitted curve, $M_{composite}(i)$, captures the nuanced hardware behaviors like latency hiding via thread switching, 
which a simple weighted average of final throughput values would miss.

\para{Final Throughput Calculation.}
After deriving a composite curve $M_{\{composite,e\}}$ for each memory engine $e$, the overall system's operation throughput is limited by the slowest parallel component. 
We find the bottleneck engine by identifying the minimum throughput across all engines at their respective total intensities ($I_e = \sum_{x \in e} i_x \cdot R_x$):
\begin{equation}
\footnotesize
    \label{eq:bottleneck}
    M_{bottleneck}(I_{bottleneck}) = \min_{e} (M_{\{composite,e\}}(I_e))
\end{equation}
The final lock-free packet throughput for the entire workload, $T_{lockfree}$, is then calculated using the throughput and intensity of only this bottlenecked engine:
\begin{equation}
\footnotesize
    \label{eq:pkt_throughput}
    T_{lockfree} = \frac{M_{bottleneck}(I_{bottleneck})}{I_{bottleneck}}
\end{equation}
Figure~\ref{fig:hybrid} validates our model's high accuracy in predicting the performance of various hybrid workloads.

\newpage
\subsection{Prediction Workflow Example}
\label{subsec:prediction_example}

This appendix expands the Predictor calculation for the running NATLB example in \S\ref{subsec:synthesizer}.
Table~\ref{tab:state_parameters} gives the full primitive parameters behind the candidate layout summarized in Table~\ref{tab:natlb_candidate}, where \textit{ConnTable} is partitioned across memory tiers and the two counter blocks reside in IMEM.
The Synthesizer derives these parameters from the IR as follows. 
The Intensity ($i$) is the number of primitive memory operations in a state block's instruction set ($I_s$). 
The Relative Rate is derived from the execution patterns ($R_s$, $R_{s,f}$). 
For state blocks like \textit{DipCntr} that are not partitioned, their relative rate is simply derived from the overall block execution rate ($R_s$). 
For a partitioned state like \textit{ConnTable}, the Synthesizer uses the per-element access rates ($R_{s,f}$) to determine the proportion of accesses directed to SRAM vs. DRAM. 
The running example's SRAM/DRAM data-access rates of 0.8/0.2 are derived this way; its remap lookup is a separate operation with rate 1.0.

\noindent\textbf{Step 1: Extract and Group Primitives.} 
From the plan, we extract all primitive memory operations $\{x\}$ and their relative rates ($R_x$), 
grouping them by their shared hardware engine bottleneck as determined by Table~\ref{tab:bottleneck}.

\noindent\textbf{Step 2: Calculate Lock-Free Throughput ($T_{lockfree}$).}
\begin{enumerate}
    \item For each engine handling multiple operations, we derive a composite performance curve, $M_{\{composite,e\}}$, using the synthesize-and-fit method described above. 
    This yields a set of composite parameters ($a_{comp}$, $b_{comp}$, $M_{\{roof,comp\}}$) for each engine.
    \item We then find the overall system bottleneck by applying the \textit{min} function across all parallel composite curves.
    \item We calculate the final packet throughput $T_{lockfree}$ using Equation~\ref{eq:pkt_throughput}. 
    This result answers Q1 and provides the parameter $a_{total}$ (from the system's final bottlenecked curve) needed for the lock model.
\end{enumerate}

\noindent\textbf{Step 3: Calculate Locked-State Throughput ($T_{lock\_max}$).}
\begin{enumerate}
    \item \textbf{Identify Locked State:} From the IR, identify the locked state block, $\{(\textit{MinLoad}, \textit{MinDip})\}$.
    \item \textbf{Isolate Critical Operations:} Extract only the memory operations performed within this block's critical section (the read and write ops from Table~\ref{tab:state_parameters}).
    \item \textbf{Derive Lock Curve:} Apply the same composition logic from Step 2 to \textit{only these critical section operations} to derive a new hypothetical performance curve. 
    From this curve, we extract the parameter $a_{lock}$.
    \item \textbf{Calculate $N_{critical}$:} Use Equation~\ref{eq:ncritical} with the $a_{total}$ (from Step 2) and $a_{lock}$ (from Step 3.3) parameters.
    \item \textbf{Calculate $T_{lock\_max}$:} Finally, calculate the maximum sustainable throughput for the locked state using Equation~\ref{eq:lock_max_throughput}. 
    This result answers Q2.
\end{enumerate}

\noindent\textbf{Step 4: Return Results.} 
The Predictor returns the final values ($T_{lockfree}$, $T_{lock\_max}$, and $N_{critical}$) to the Synthesizer, which uses them to validate the allocation plan.

\section{Flow Size Distribution}

{
\setlength{\abovecaptionskip}{4pt}
\begin{figure}[tp] \centering
\includegraphics[width=0.6\columnwidth]{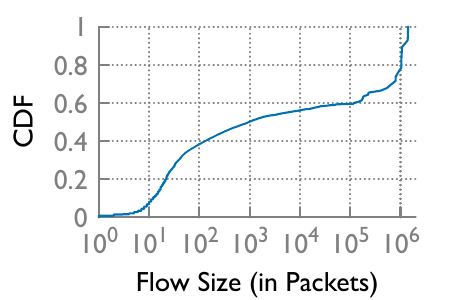}
\caption{Flow size distribution of CIC-IDS2017 trace.}
\label{fig:fb}
\end{figure}
}

Figure~\ref{fig:fb} shows the flow size distribution of CIC-IDS2017 network traces.